\documentclass{article}

\newtheorem{lemma}{Lemma}
\newtheorem{proposition}{Proposition}

\newcommand{\ux}{\textsc{$U3$}\xspace}
\newcommand{\uh}{\textsc{$U2$}\xspace}
\newcommand{\uz}{\textsc{$U1$}\xspace}

\newcommand{\bra}[1]{\langle #1|}
\newcommand{\ket}[1]{|#1\rangle}

\usepackage{xspace}
\usepackage{color}
{}
{}
{}
\usepackage{url}
\usepackage{multirow}
\usepackage{enumerate}
\usepackage{graphicx}
\usepackage{braket}
\usepackage{todonotes}
\usepackage{amsmath}
\usepackage{hyperref}
\usepackage{amsmath}   
\usepackage{mathrsfs}
\usepackage{amsfonts}
\usepackage{mathtools}
\usepackage{comment}
\usepackage{longtable}
\usepackage{multicol}
\DeclarePairedDelimiter\ceil{\lceil}{\rceil}
\usepackage{wrapfig}
\usepackage{lscape}
\usepackage{hyperref}
\usepackage{rotating}
\usepackage[ruled,vlined]{algorithm2e}
\DeclareUnicodeCharacter{2212}{-}

\usepackage{mathtools}
\usepackage{PRIMEarxiv}

\usepackage[utf8]{inputenc} 
\usepackage[T1]{fontenc}    
\usepackage{hyperref}       
\usepackage{url}            
\usepackage{booktabs}       
\usepackage{amsfonts}       
\usepackage{nicefrac}       
\usepackage{microtype}      
\usepackage{lipsum}
\usepackage{fancyhdr}       
\usepackage{graphicx}       
\graphicspath{{media/}}     

\title{Circuit Design for Clique Problem and Its Implementation on Quantum Computer
}

\author{Arpita Sanyal (Bhaduri)$^1$, Amit Saha$^{1,3*}$, Debasri Saha$^{1}$, Banani Saha$^2$, Amlan Chakrabarti$^{1}$\\~\\

$^1$A. K. Choudhury School of Information Technology, University of Calcutta\\
$^2$Computer Science and Engineering, University of Calcutta\\
$^3$Atos, Pune, India\\~\\
$^*$abamitsaha@gmail.com}

\begin{document}
\maketitle

\begin{abstract}
 Finding cliques in a graph has several applications for its pattern matching ability. \textit{k}-clique problem, a special case of clique problem, determines whether an arbitrary graph contains a clique of size \textit{k}, has already been addressed in quantum domain. A variant of $k$-clique problem that lists all cliques of size $k$, has also popular modern-day applications. Albeit, the implementation of such variant of $k$-clique problem in quantum setting still remains untouched. In this paper, apart from theoretical solution of such  \textit{k}-clique problem, practical quantum gate-based implementation has been addressed using Grover's algorithm. This approach is further extended to design circuit for the maximum clique problem in classical-quantum hybrid architecture. The algorithm automatically generates the circuit for any given undirected and unweighted graph and any given $k$, which makes our approach generalized in nature. The proposed approach of solving $k$-clique problem has exhibited a reduction of qubit cost and circuit depth as compared to the state-of-the-art approach, for a small $k$ with respect to a large graph. A framework that can map the automated generated circuit for clique problem to quantum devices is also proposed. An analysis of the experimental results is demonstrated using IBM's Qiskit.
\end{abstract}


\section{Introduction}

Quantum computers were proposed in the early 1980s and the description of quantum mechanical computers was formalized in the late 1980s. Many efforts on quantum computers have progressed steadily since the early 1990s, because these computers were shown to be more powerful than classical computers on various specialized problems, especially on computationally NP-problems. Several quantum algorithms, for example \textbf{Shor's Algorithm} \cite{31} for factoring integers, \textbf{Grover's Algorithm} \cite{32} for searching an unstructured database, \textbf{Triangle finding} by Magniez et al. \cite{33}, \textbf{Matrix Product Verification} \cite{34} have already been proposed and shown asymptotic improvements than their classical counterparts. In this paper, another computationally NP-problem $i.e.,$ clique problem \cite{41} has been addressed in quantum setting. Our main focus in this paper is to provide an end-to-end framework that automatically implements an clique problem, so that if anyone can map their computational problem to the clique problem in polynomial time, will be able to implement further, without prior knowledge of gate-based quantum circuit implementation.    

A clique is a subgraph of an undirected graph, where every distinct vertex in the subgraph is connected with every other vertex via an edge; that is, the subgraph is complete. The \textit{k}-clique problem is a special case of clique problem which determines whether an arbitrary graph contains a clique of size \textit{k}. Whereas, a maximum clique is a complete sub graph of a graph, whose size is the largest among all other complete sub graphs in the given graph. Clique problems have several applications in various branches of computer science, as for instance, pattern recognition \cite{8}, information retrieval \cite{1}, computer vision \cite{8}, analysis of financial networks \cite{9} and  spatial data mining \cite{5}. Solving the clique problem especially $k$-clique problem using quantum algorithm \cite{sota}, where the solution provides a clique of size $k$, is more efficient in terms of computation as compared to its classical counterpart due to its quantum mechanical properties. Quantum circuit design for the \textit{k}-Clique problem has also been demonstrated in the literature \cite{sota}. Although, circuit designing in quantum setting for a variant of $k$-clique problem, where the solution of the problem lists all cliques of size $k$ is much more difficult in reality due to the computational complexity of this variant of $k$-clique problem.
 
 In this paper, we solve the variant of $k$-clique problem, which lists all the cliques of size $k$ using well-known quantum search algorithm $i.e.,$ Grover's algorithm to get the immediate advantage of solving various modern-day applications like community detection \cite{2,3,4}, data mining in bio-informatics \cite{6} and disease classification \cite{7}. We have designed the circuit for specified variant of $k$-clique problem in such a way, so that the engineering challenge of implementing such variant of $k$-clique problem must be overcome. We further illustrate that for small value of $k$ with comparatively larger graph, our approach of designing circuit for $k$-clique problem surpasses the state-of-the-art approach \cite{sota} with respect to the quantum cost with regards to qubit and circuit depth. Using this proposed approach of solving $k$-clique problem, a generalized algorithm for the maximum clique problem (MCP), which lists all the largest sized cliques among all the cliques for a given graph has also been proposed in this paper.
 
 Our key contributions in this paper can be summarized as follows: 
\begin{itemize}

\item We propose an automated end-to-end framework for mapping clique problem to any available quantum computer so that anyone can get the advantage of gate-based implementation of quantum algorithm without much prior knowledge.

\item We propose an approach to implement the variant of $k$-clique problem, where the output of the problem lists all cliques of size $k$ using quantum search algorithm for the first time to the best of our knowledge.

\item Our approach of solving $k$-clique problem outperforms the state-of-the-art approach with respect to qubit cost and circuit depth, when $n >> k$, where $n$ is a large number of nodes of given graph ($G(V=n, E=e)$) and $k$ is respectively very small.

\item We exemplify the triangle finding problem, which is an example of $k$-clique problem, where $k=3$ to establish our claim.

\item Further, we extend our approach of solving $k$-clique problem to implement maximum clique problem via classical-quantum hybrid computation.

\item We implement the generalized algorithm on arbitrary graph instances and simulated the same in the QASM simulator as well as in a real quantum devices (IBMQX architecture \cite{27}, \cite{28}) through a python based programming interface called QISKit \cite{29} with noise model and study that for different graphs, how the error effects the resultant states further.
\end{itemize}

The paper has been organized as follows. \textbf{Section 2} reviews related works in the classical and quantum domain. The general flow of the automated framework for clique problem is illustrated in \textbf{Section 3}. Our proposed algorithm for synthesis of the \textit{k}-clique problem has been discussed in \textbf{Section 4}. The extension of the algorithm for solving MCP has also been depicted in this section. The performance of the circuit is analyzed in \textbf{Section 5}. \textbf{Section 6} deals with experimental results of the circuit genrated for different exemplified graphs in the $IBMQ\_qasm\_simulator$ and $IBMQ\_16\_melbourne$. The complexity of the algorithm has been analyzed in \textbf{Section 7}. Finally the paper concludes with a summary and future scope in \textbf{Section 8}.

\section{Background}
In this section, we will try to lay out some background knowledge
on Quantum computing and clique problem.

\subsection{Qubits and Quantum States} The qubit is the quantum equivalent of a classic bit \cite{chuang}. Instead of working with classical bits, quantum computers, or more precisely quantum processing units (QPUs), use qubits. Qubits are mathematically expressed using dirac notation. This notation is also known as bra-ket notation. An example of complete bra-ket notation is
$\bra{\phi}\ket{\psi}$, where $\phi$ represents ket part and $\psi$ represents bra part. A  bit is a binary information that can have two possible values: 0 or 1. On the contrary, a qubit can be expressed as $\ket{\psi}=\alpha\ket{0}+\beta\ket{1}$, where $\alpha$ and $\beta$ are amplitudes of the state $\ket{0}$ and $\ket{1}$ and $\alpha^2+\beta^2=1$.  The states $\ket{0}$ and $\ket{1}$ are the basis states, similar to bit values of 0 and 1 in classical computers. To denote a quantum state, only the ket part of the notation is used. For example, to represent the classical values of 0 and 1 in a quantum state, we can write as
$\ket{0}$=
$\begin{pmatrix}
 1 \\ 0
\end{pmatrix}$, $\ket{1}=$
$\begin{pmatrix}
 0 \\ 1
\end{pmatrix}$.
The column vector represents amplitude of the quantum state.

\subsection{Quantum Logic Gates}
A quantum circuit is a model for quantum computation, where reversible quantum gates like Hadamard, CNOT, Toffoli are imposed on qubits to evolve the quantum states towards solution of a specific quantum algorithm \cite{PhysRevA.52.3457}. Quantum gates are the unitary operations, where number of gate collections to be sequentially executed defines the depth of a quantum circuit. Whereas, The number of qubits defines the width of the circuit. The matrix representation of the quantum gates that are used in the proposed approach is shown in the following Table \ref{tab:qg}.
\begin{table}[ht!]
\centering
\caption{Matrix Representation of Quantum Gates}
\scriptsize
\begin{tabular}{  c | c | c }
 \hline
 Quantum Operator & Quantum gate & Matrix Representation \\
 \hline
 Hadamard & \includegraphics[scale=.5]{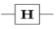} & $\begin{pmatrix}
\\ \frac{1}{\sqrt{2}} & \frac{1}{\sqrt{2}}
\\ \frac{1}{\sqrt{2}} & -\frac{1}{\sqrt{2}}
\end{pmatrix}$ \\
 \hline
 NOT & \includegraphics[scale=.3]{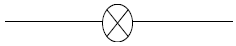} &  $\begin{pmatrix}
\\ 0 & 1
\\ 1 & 0
\end{pmatrix}$\\
 \hline
CNOT & \includegraphics[scale=.5]{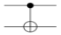} & $\begin{pmatrix}
\\ 1 & 0 & 0 & 0
\\ 0 & 1 & 0 & 0
\\ 0 & 0 & 0 & 1
\\ 0 & 0 & 1 & 0
\end{pmatrix}$ \\
\hline
 TOFFFOLI & \includegraphics[scale=.5]{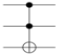} &  $\begin{pmatrix}
\\ 1 & 0 & 0 & 0 & 0 & 0 & 0 & 0
\\ 0 & 1 & 0 & 0 & 0 & 0 & 0 & 0
\\ 0 & 0 & 1 & 0 & 0 & 0 & 0 & 0
\\ 0 & 0 & 0 & 1 & 0 & 0 & 0 & 0
\\ 0 & 0 & 0 & 0 & 1 & 0 & 0 & 0
\\ 0 & 0 & 0 & 0 & 0 & 1 & 0 & 0
\\ 0 & 0 & 0 & 0 & 0 & 0 & 0 & 1
\\ 0 & 0 & 0 & 0 & 0 & 0 & 1 & 0
\end{pmatrix}$ \\
 \hline
\end{tabular}
\label{tab:qg}
\end{table}

\begin{figure}[ht!]
\centering
\includegraphics[width= 3in]{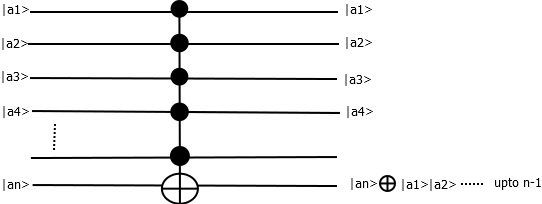}
\caption{Multi-Control Toffoli gate} 
\label{mct}
\end{figure}

Apart from these four gates described in Table \ref{tab:qg}, Multi-Controlled Toffoli Gate (MCT) is also used to design our proposed circuit for clique problem. There are \textit{$n$} number of inputs and outputs in an $n$-qubit MCT. This MCT gate passes the first $n-1$ inputs, which are referred as control bits to the output unaltered. It inverts the $n^{th}$ input, which is referred as the target bit if the first $n-1$ inputs are all one. An MCT gate is shown in Figure \ref{mct}, where black dots $\bullet$ represent the control bits and the target bit is denoted by a $\oplus$.

\subsection{NISQ Devices}
\begin{figure}[ht!]
\centering
\includegraphics[width=50mm, height=2cm]{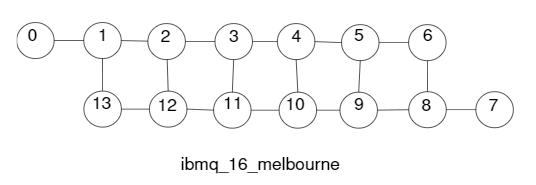}
\caption{Qubit Topology \cite{43}}
\label{qt}
\end{figure}

In recent years, NISQ (Noisy Intermediate Scale Quantum devices) \cite{Preskill2018quantumcomputingin} that can perform quantum computation with a short circuit length have appeared, although the scale and accuracy are insufficient to perform continuous and effective error correction. Various physical systems such as superconductors, ion traps, quantum dots, NV centres, and optics are used in NISQ devices. As an instance, superconducting based IBM Q processors accept gates written in the QASM language \cite{45}, which is further used in this paper for all the experiments that we have performed. Specifically, we have used IBM's Melbourne quantum device. The qubit topology of IBM's Melbourne is as shown in Figure \ref{qt}.

All multi-controlled logical gates of the generated logical circuits are needed to be decomposed into four types of one-qubit and two-qubit gates to map into IBM Q processor. We describe those gate sets and the required pulses in the IBM Q superconducting processors in Table \ref{tab:gates}. Since no pulse is required, we can perform  $\uz$~with zero cost. The error level of $\ux$~is twice of $\uh$~and approximately an order of magnitude less than the $CX$ gate.
\begin{table}[ht!]
\centering
\caption{Gate Set for QASM}
\begin{tabular}{c|c}
gate type &  remarks \\ \hline
$U1(\lambda)$      & No pulse.     Rotation $Z$~($R_Z$) gate.   \\
$U2(\phi,\lambda)$    & One $\frac{\pi}{2}$ pulse.     $H$ gate is $U2(0,\pi)$. \\
$U3(\theta, \phi, \lambda)$     & Two  $\frac{\pi}{2}$ pulses.   $R_Y(\theta)$ gate is $U3(\theta, 0, 0)$. \\
$CX$ &  Cross-resonance pulses and One $\frac{\pi}{2}$ pulse.
\end{tabular}
\label{tab:gates}
\end{table}

\begin{figure}[ht!]
\centering
\includegraphics[width=80mm, height=4cm]{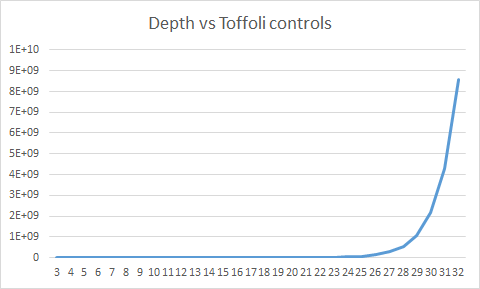}
\caption{Circuit depth vs Toffoli controls on IBM Q: Toffoli controls along with x axis and depth along with y axis}
\label{dvtc}
\end{figure}

In Figure \ref{dvtc}, we have depicted that the circuit depth increases with respect to the number of Toffoli controls through implementation of MCT gate on IBM Q device. With the increase of depth, error in quantum circuit also becomes colossal, which is thoroughly described next.

\subsubsection{Errors in NISQ Devices}
Qubits are not stable as even a small perturbation in the environment can change the state of a qubit. Error rate for a qubit can be defined as probability of undesired change in the qubit state. Errors in quantum computers can be classified into two categories: retention-errors and operational-errors.

\textbf{Retention Errors or Coherence Errors:}
A qubit can retain data for only a limited time, and this duration is called as Coherence Time. There are two types of retention errors that can occur, and there are two metrics to specify the coherence time of a quantum device. A qubit in an high-energy state (state $\ket{1}$) naturally decays to the low-energy state (state $\ket{0}$), and the time constant associated with this decay is called $T1$ Coherence Time. However, there is also a possibility that qubit might interact with environment and encounter a phase error even before relaxing into $\ket{0}$ state, and the time constant associated with this decay is called $T2$ Coherence Time. $T2$ indicates the time for a qubit to get affected by the
environment.

\textbf{Operational Errors or Gate Errors:} Quantum operations do not give perfect result. So operations on qubits can also affect their state incorrectly due to errors, for example, an instruction that rotates the state by some desired angle can introduce extra erroneous rotation. Operational error-rate is defined as the probability of introducing an error while performing the operation. Thus, if the depth of a circuit is high, operational errors become huge.

\subsection{General Hybrid Architecture for Quantum Algorithms}
\begin{figure}[ht!]
\centering
\includegraphics[width=3in]{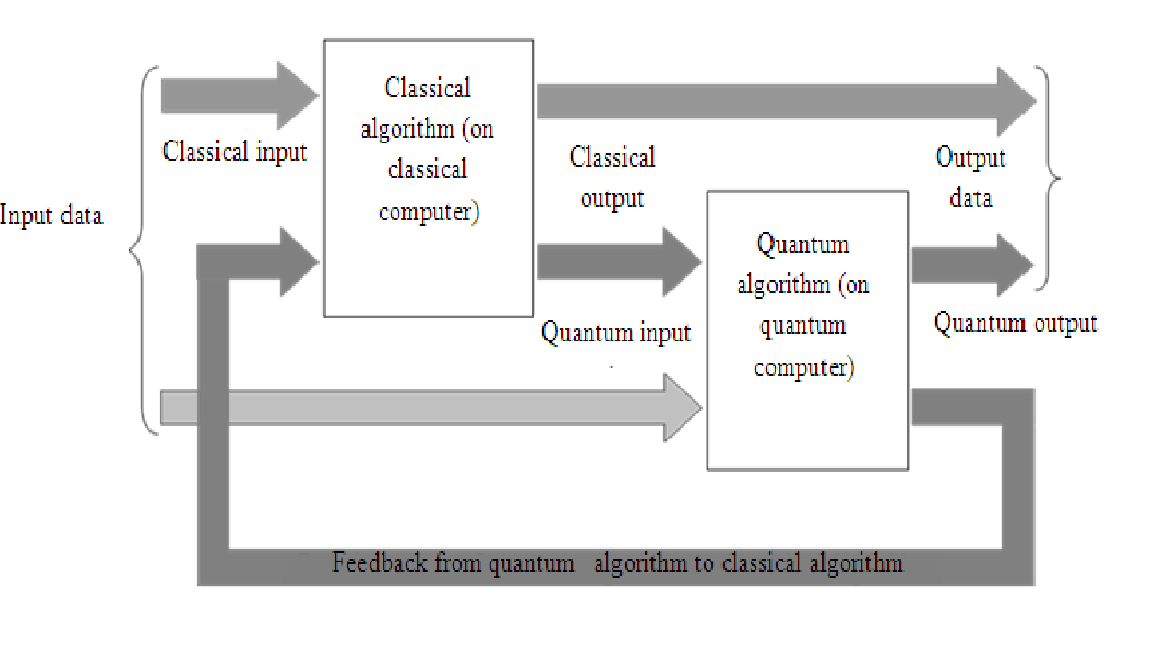}
\caption{The relationship between classical part and quantum part of the hybrid algorithm \cite{17}}
\label{hybrid}
\end{figure}

In this paper, the concept of hybrid quantum and classical computing is used \cite{17} to solve maximum clique problem. In a hybrid quantum classical architecture, a classical algorithm is used for performing some basic operations(preparation of inputs, feeding output to the quantum machine) which may be required by the quantum algorithm. The quantum subsystem of this architecture initializes the quantum registers/nodes, prepares the inputs by applying quantum gates, execute the quantum oracle with the help of quantum gates and unitary transformations. Finally, evaluate the result stored in the quantum registers and send them to classical computer as feedback for further processing. The general architecture of the hybrid model is shown in Figure \ref{hybrid}. Next, let us put some light on the fundamental concepts of Grover's algorithm and clique problem, which are of utmost importance with respect to our proposed approach.

\subsection{Grover's Algorithm} 
\begin{figure}[ht!]
\centering
\includegraphics[width=80mm,height=4.0cm]{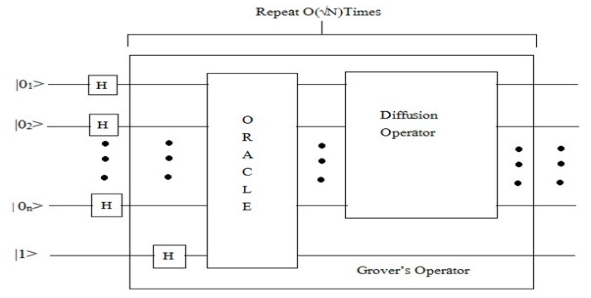}
\caption{Generalized circuit for Grover's algorithm \cite{32}}
\label{grover}
\end{figure}

Grover's algorithm\cite{32} is a way of finding an element in an unsorted list with $N$ elements using quantum computers. Grover's algorithm is based on amplitude amplification of the basis state which specifies a position of a searched element in the list. It runs in time O$(\sqrt{N})$, where $N$ is the number of elements in the list. The generalized structure of Grover's algorithm is shown in Figure \ref{grover}.

\par Grover’s algorithm works with a unitary operator $O$ called the oracle function, which is defined by $\ket {x} \ket{q} \implies \ket{x} \ket {q \oplus f(x)}$. where $\ket{x}$ is the $n$ qubit index register and $\ket{q}$ is an additional qubit, called the oracle qubit. The functional view of Grover's Search Algorithm is presented here. 
 $\exists$ function oracle $(O)$ such that

\[
   						 O\ket{x}= 
							\begin{cases}
    							-\ket{x},& \text{if \textit{x} is Marked}\\
                                 \ket{x},              & \text{otherwise}
\end{cases}
\]

\par More elaborately, The steps of the Grover's algorithm are as follows:
 
\par \textbf{Initialization}: The algorithm starts with the uniform superposition of all the basis states on input qubits $n$. The last ancilla qubit is used as an output qubit which is initialized to $H \ket1$. Thus, we obtain the quantum state $\ket \psi$.

\textbf{Sign Flip}: Flip the sign of the vectors for which the oracle gives output 1.

\textbf{Amplitude Amplification}: We need to perform the inversion about the average of all amplitudes of quantum state for a certain number of iterations to make the amplitude of the marked state large enough, so that it can be obtained from a measurement with probability close to 1. This phenomenon is known as amplitude amplification which is performed by using a diffusion operator. 

\textbf{Number of Iterations}: Iterations of Grover’s algorithm is the number of times that the oracle and amplification stages are performed.
Each iteration of the algorithm increases the amplitude of the marked state by $O(\sqrt{1/N})$. In this way, Grover's search algorithm requires $\sqrt{N/M}$  iterations to get the probability of one of the marked states $M$ out of total $N$ number of states set. If the number of iterations of Grover's algorithm are more than the optimal number, the probability of measuring the desired state actually goes down.
Therefore, right number of iterations are important for getting proper result. 
\par{\textbf{Diffusion operator}}:
This diffusion operator of Grover's algorithm \cite{32} inverts the amplitude of the input states about their mean value of amplitude. The generalized matrix representation of the diffusion operator is shown in Table \ref{diffusion}. A six-qubit diffusion operator is also presented in Table \ref{diffusion}. As shown in six-qubit diffusion operator, a 6-qubit Toffoli gate is required. Therefore, for $n$-qubit diffusion operator, $n$-qubit Toffoli gate is needed. This $n$-qubit Toffoli gate needs to be realized into one-qubit or two-qubit gates as discussed. While decomposing the $n$-qubit Toffoli gate, if the depth and the ancilla qubits increase arbitrarily then the time complexity of the algorithm also increases, which is undesirable.
\begin{table}[htb]
\centering
\caption{Diffusion Operator}
\begin{tabular}{ c | c }
\hline
 Six-qubit Diffusion Circuit & Generalized Matrix Representation \\ \hline
\\ \includegraphics[width=3cm]{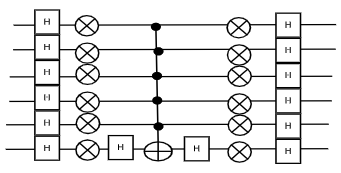} &
$\begin{pmatrix}
 -1 + \frac{2}{N} & \frac{2}{N} & \ldots & \frac{2}{N}\\ 
 \frac{2}{N} & -1 + \frac{2}{N} & \ldots & \frac{2}{N}\\
\vdots & \vdots  & \ddots & \vdots\\ 
\frac{2}{N} & \frac{2}{N} & \ldots &  -1 + \frac{2}{N}
\end{pmatrix}$ \\
\\ \hline
\end{tabular}
\label{diffusion}
\end{table}

\subsection{Clique Problem}

 This subsection deals with the definition of a clique, \textit{k}-clique, maximum clique of a graph and the clique problem in quantum domain, which are used in the rest of the paper. \par
\subsubsection{Clique, \textit{k}-clique and maximum clique} 
\begin{figure}[ht!]
\centering
\includegraphics[width=1in]{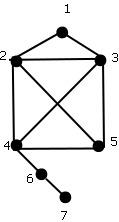}
\caption{An arbitrary graph}
\label{arbitrarygraph}
\end{figure}

A clique is a complete sub graph of a graph. Particularly, if there is a subset of \textit{k} vertices that are connected to each other, we state that the graph contains a \textit{k}-clique. Let \textit{$G(V,E)$} be a graph, where \textit{$V$} be the set of vertices and \textit{$E$} be the set of edges. If \textit{${u,v} \in E$}, then \textit{u} and \textit{v} are said to be adjacent. The set of vertices adjacent to a vertex \textit{$v$} is called the neighbourhood of \textit{$v$} and is denoted by \textit{$N(v)$}.  A clique of a graph \textit{G} is a set of vertices \textit{$C$} in which \textit{${u, v} \in C \implies {u, v} \in E$}. We say that the graph contains a \textit{k}-clique, if there is a subset of \textit{k} vertices that are connected to each other. A maximum clique is a complete sub graph of a graph \textit{G}, whose size is largest among all other complete sub graphs in \textit{G}. In the graph of Figure \ref{arbitrarygraph}, there are six vertices, but available cliques are $(2 3 4) (3 5 2) (5 4 3) (4 2 5) (2 3 4 5)$ and maximum clique is $(2 3 4 5)$.
 
 \subsubsection{Clique Problem in Quantum Domain} The most commonly studied clique problem specifically \textit{k}-clique problem is the 3-clique problem or triangle finding problem. It is quite trivial to show that the randomized classical query complexity of the \textit{3}-clique problem is $O(n^2)$ where \textit{n} denotes the number of vertices in the graph \cite{40}. Szegedy constructed a quantum algorithm for the \textit{3}-clique problem with query complexity $O(n^\frac{10}{7})$ \cite{37}. Later, Magniez et al. showed, that the \textit{3}-clique problem can be solved with improved query complexity $O(n^\frac{13}{10})$ using a quantum walk approach \cite{33}. To date, the best known lower bound on the quantum query complexity of the \textit{3}-clique problem is $\Omega(n)$ and the best known algorithm has a complexity of $O(n^\frac{5}{4})$ \cite{38} \cite{39}. Further, using quantum algorithms, some research works have also been done for the $k$-clique problem, when $k > 3$ \cite{33, sota}. Among all these works on $k$-clique problem, the state-of-the-art work \cite{sota} showed that to implement $k$-clique problem, at least $n$ number of input qubits are required, when the number of vertices of given graph are $n$. In addition total $log_2 k + 3$ ancilla qubits are required to perform Grover's algorithm for solving $k$-clique problem. In this paper, we compare our proposed work with this state-of-the-art work on $k$-clique problem. 
 
 We also propose a generalized approach of solving MCP with the help of proposed approach of $k$-clique problem. There are some existing works on MCP in quantum computing, which are needed to be discussed. In 2015, Pronaya Prosun Das et al. \cite{15} used the concept of quantum inspired evolutionary algorithm  proposed by Kuk-Hyun Han et al \cite{14} for solving the MCP. A quantum inspired evaluation algorithm is a combination of quantum computing and evolutionary algorithm. Elijah Pelofske et al. proposed quantum annealing for solving the MCP \cite{16}. With this background, we are inclined to contribute some research advancement on clique problem in quantum domain. 

\section{General Flow of Proposed Automated Framework for Mapping Clique Problem to Quantum Computers}
\begin{figure}[ht!]
\centering
\includegraphics[width= 4in]{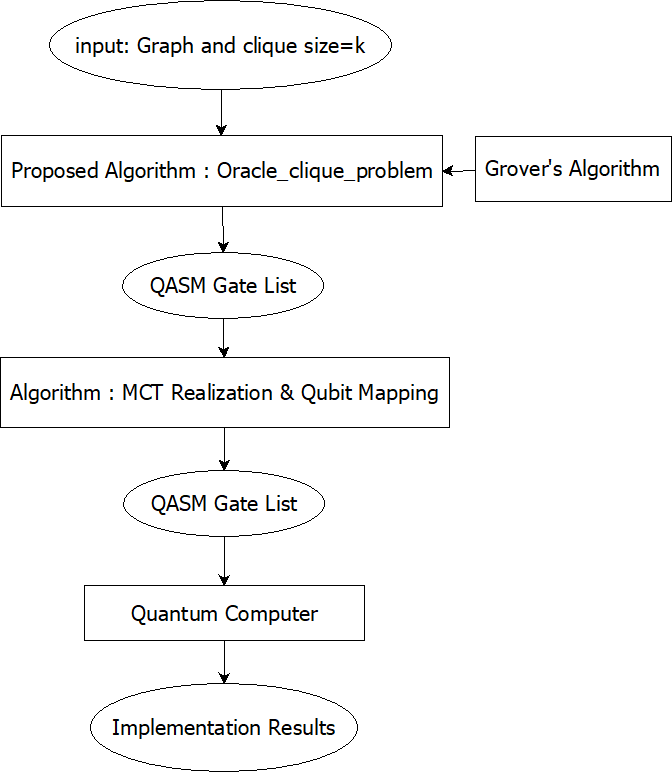}
\caption{The complete flow of proposed work}
\label{generalflow}
\end{figure}

The complete flow of proposed design for mapping clique problem from a graph to quantum computer is shown in Figure \ref{generalflow}. The proposed algorithm \textbf{Oracle\_clique\_problem} takes adjacency matrix as input and gives an QASM gate list as output. After realizing the logical gates into quantum computer supported 1-qubit and 2-qubit gates and through qubit mapping algorithm based on qubit topology, this QASM gate list can be further used as an input to a quantum computer to get the implementation result of clique problem. These MCT realization \cite{Acasiete_2020} and qubit mapping algorithm \cite{zul8342181, li2019tackling, tan2020optimal} are well-defined in the literature. We have just adopted them as a support in our framework.

The main contribution that we have made in this paper is to automatically design the circuit for clique problem using Grover's algorithm, when the adjacency matrix of graph and clique size $k$ are given. As discussed in subsection 2.5, circuit design of diffusion operator is generalized irrespective of the given computational problem. Moreover, the oracle is problem specific, therefore our proposed algorithm is completely based on the generation of oracle circuit for clique problem.

\section{Proposed Methodology of Circuit Design for Clique Problem}
This section outlines the proposed methodology for the circuit synthesis of the \textit{k}-clique problem using Grover’s search algorithm. \textit{k}-clique problem takes an adjacency matrix of a graph as input and determines the cliques of size \textit{k} exist within the graph. As mentioned in the previous section, there are two parts of the Grover’s search algorithm. While the diffusion operator is predefined for all problem instances, the oracle is specific to the given search problem. Hence, this paper proposes the design of an oracle circuit that takes a graph as input and determines if the graph contains a clique of size \textit{k}. The oracle then marks those states that form a clique before the diffusion operator is applied to them. This proposed approach of $k$-clique problem is further used to solve MCP later in this section. Let's start with the proposed oracle for $k$-clique problem.

\subsection{Proposed Oracle for \textit{k}-Clique Problem} 
 The aim of this paper is to construct the quantum circuit block for \textit{k}-clique problem. Theoretically, the oracle is only a function that checks whether a specific item is the target or not. However, due to the linearity of quantum mechanics, when the oracle is applied to the superposition state $\ket{\psi_{0}}$, all possible items are examined against the criteria. 
 In this generalized circuit, there are seven main steps, which are Initialization, Hadamard transformation, Qubit activation, Edge detection, Clique detection, Qubit deactivation and Phase flip. For finding \textit{k}-clique in a graph, all the steps have been described in the following subsections.


\subsubsection{Initialization}
If there are $n$ vertices in the input graph, then the number of qubits required to represent each vertex is $\ceil{\log_2 n}$. The oracle checks a combination of \textit{k} vertices from a combination of all \textit{k} vertices without duplicate vertices set at a time to determine if a clique is formed by them. Hence, a total of $m=k*\ceil{\log_2 n}$ input qubit lines are required to represent a combination of \textit{k} vertices for input. Total number of ancilla qubits required to verify whether the subgraph formed by \textit{k} number of vertices is complete or not is $k \choose 2$ $+1$. The initial input qubits include $m$ qubits prepared in the ground state $\vert\psi\rangle=\ket{0}^{\otimes m}$, ({$k \choose 2$} +1) ancilla qubits in the ground state $\ket{\theta}=\ket{0}^{\otimes {k \choose 2} +1}$ (These ($k \choose 2$ +1) ancilla qubits are required to prepare clique detector block, which is described in the next subsection thoroughly) and one output qubit in the excited state $\ket{\phi}=\ket{1}$,  which is required to perform the CNOT operation of the oracle. This entire initialization can be mathematically  written as:

\[\vert\psi\rangle\otimes\vert\theta\rangle\otimes\vert\phi\rangle=\vert 0\rangle^{\otimes m}\otimes\vert 0\rangle^{\otimes {k \choose 2} +1}\otimes\vert 1\rangle\]
\subsubsection{Hadamard Transformation}

After the initialization, the Hadamard transform $H^{\otimes m}$ on input qubits and $H$ on output qubit is performed, therefore all possible states are superposed as $\ket{\psi_{0}}\otimes\ket{\theta_{0}}\otimes\ket{\phi_{0}}$, where

\[\vert\psi_{0}\rangle=\,{1\over\sqrt{2^{m}}}\sum_{i=0}^{2^{m}-1}\vert i\rangle\]

\[\ket{\theta_{0}}=\ket{0000\dots0}\]

\[\vert\phi_{0}\rangle=\,{1\over\sqrt{2}}\left(\vert 0\rangle-\vert 1\rangle\right)\]

 \subsubsection{Qubit Activation}

 For a given graph, vertices need to be numbered as $\{0,1,2 \dots n-1\}$. The input qubit lines act as the binary representation of combination of \textit{k} vertices. After Hadamard transformation on input qubit lines, $\ket{\psi_{0}}$ represents all possible combination of \textit{k} vertices of given graph. But, the oracle checks only the combination of \textit{k} vertices without any duplicate entry of vertices to determine if a clique exists or not. After Hadamard transformation, we get superposition of all the states which includes several invalid states. So we have reduced the search space by excluding all the invalid vertex combinations that do not contribute to clique. Only valid vertex combinations that may form a clique are considered using a technique named as Qubit Activation. To make sure that the oracle is checking only all possible combinations of \textit{k} vertices without duplicate entry, all the input qubit lines are needed to be in the excited state $\ket{1}$ for those particular combination of \textit{k} vertices to make the input qubit lines suitable as control lines for Multi Control Toffoli (MCT) operation. A number of NOT gate has to be imposed on the input qubit lines, which are in the ground state $\ket{0}$ for every possible combination of $k$ vertices without duplicate entry. 
 The qubit activation block activates a qubit by applying NOT gate if it is 0 to make the desired inputs are string of 1's.  

 \subsubsection{Clique Detector Block}

\begin{figure}[ht!]
\centering
\includegraphics[width=10cm, height=6cm]{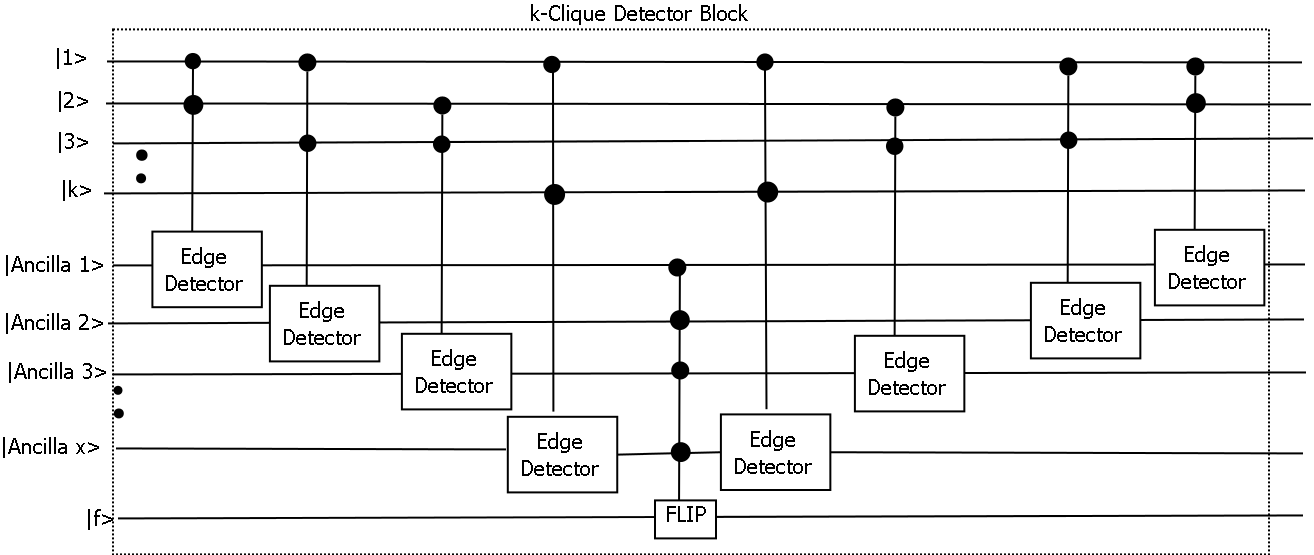}
\caption{Circuit representation of k-clique Detector Block}
\label{cliquedetector}
\end{figure}

The proposed Clique Detector Block is shown in Figure \ref{cliquedetector}. This block is defined as follows:\\
\[
  CliqueDetector(v_1, v_2, v_3, \dots, v_k, f) =
  \begin{cases}
                                   \text{$f=1$} & \text{if $v_1$, $v_2$, $v_3$ \dots $v_k$ form a clique} \\
                                   \text{$f=0$} & \text{otherwise} \\
  \end{cases}
\]
where $v_1$, $v_2$, $v_3$, $\dots$, $v_k$ are the combination of \textit{k} vertices of the input graph, which are activated input qubit lines and the $f$ is the circuit output of Clique Detector Block, which is represented by $({k \choose 2} +1)^{st}$ ancilla qubit line. Then, the Edge Detector Block imposes if a pair of vertices of input graph is adjacent. This edge detecting sub-circuit is defined next.

\subsubsection{Edge Detector Block} The function of the Edge Detector Block is to detect edges between pair of vertices. A Multi-controlled Toffoli gate is applied for checking the connectivity between a pair of vertices. One MCT gate indicates presence of edge between a pair of vertices. So for detecting multiple edges, multiple MCT gates are needed. 

\[
  EdgeDetector(v_1,v_2,f) =
  \begin{cases}
                                   \text{$f=1$} & \text{if $v_1$ and $v_2$ are adjacent} \\
                                   \text{$f=0$} & \text{otherwise} \\
  \end{cases}
\]
This sub-circuit Edge Detector Block checks if two vertices are adjacent and changes the state of the output line to $\ket{1}$. First $k \choose 2$ ancilla lines are the representation of these output lines. If this is true to all $k \choose 2$ ancilla lines, the Clique Detector confirms the presence of a clique in the form of $\ket{1}$ state on the $({k \choose 2 }+1)^{st}$ ancilla line, otherwise the state of $({k \choose 2} +1)^{st}$ ancilla line remain same as it's initial state. Therefore, if there exists a clique, the ancila qubit state becomes $\ket{\theta_{1}}$ for respective input vertices. \[\ket{\theta_{1}}=\ket{1111\dots1}\]
Clique detection block detects a clique by applying Toffoli gate between the output of edge detection blocks. 

\subsubsection{Qubit Deactivation} A number of NOT gate has to be imposed on the input qubit lines in the reverse order to deactivate the qubits again or to reset the input qubits to their initial value. \par  
\subsubsection{CNOT Operation}
\begin{figure*}[ht!]
\centering
\includegraphics[width=16cm,height=5.5cm]{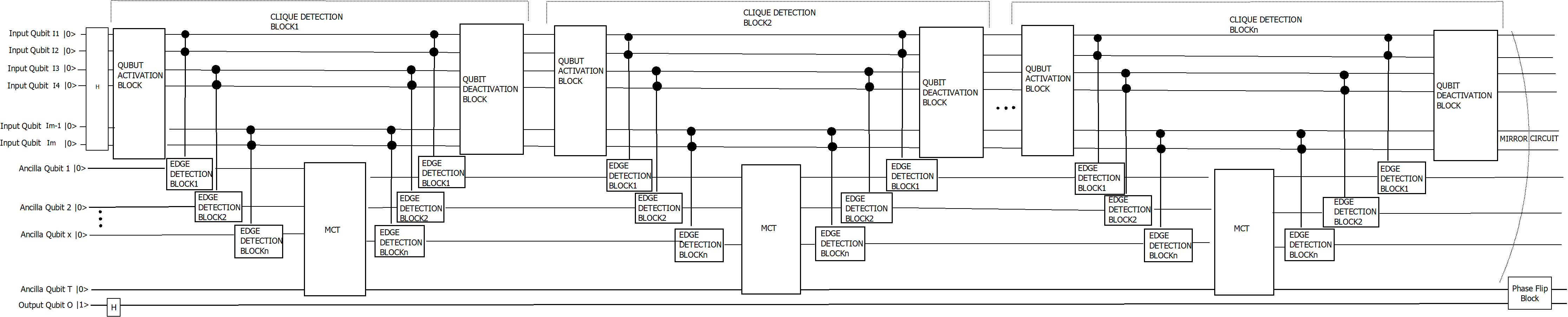}
\caption{Circuit representation of oracle for \textit{k}-clique problem}
\label{kclique}
\end{figure*}

The output qubit state $\vert\phi_{0}\rangle$ is initially set as
${1\over\sqrt{2}}\left(\vert 0\rangle-\vert 1\rangle\right)$. Applying an CNOT gate on output line considering $({k \choose 2 }+1)^{st}$ ancilla qubit as control results in an eigenvalue kickback $-1$, which causes a phase shift for the respective input state, which makes a clique. The Figure \ref{kclique} shows the generalized view of the oracular circuit for k-clique problem.

\subsection{Generalized Algorithm for Solving \textit{k}-Clique Problem}
The proposed algorithm for designing \textit{k}-clique problem takes adjacency matrix as input and gives QASM gate list as output. This QASM gate list can further be used as an input to quantum simulator for verification or to map into quantum technologies. The algorithm for the synthesis of \textit{k}-clique problem has a classical control that consist of classical components, which helps to build the quantum part of the algorithm. This algorithm can be illustrated as follows:
\begin{itemize}
\item \textbf{Classical input}:  
\begin{itemize}
\item Adj[][]: Adjacency matrix of the input graph G(V, E). 
\item arr[]: Array that holds the vertices of the graph.
\item $comb_{arr}[]$: Array that holds all combinations of the vertices for a clique size.
\item active[]: Array that holds the binary equivalent of all combinations of vertices.
\end{itemize}
\item \textbf{Classical part of the algorithm}:
\begin{itemize}
\item Classical part of the algorithm prepares all possible input combination of the vertices for a clique size. It creates $n \choose k$ combination of vertices if n is the total number of vertices and k is clique size.
\end{itemize}
\item \textbf{Quantum input}:
\begin{itemize}
\item I[]: This is a quantum register that holds the input qubits.
\item A[]: This is a quantum register that holds the ancilla qubits.
\item T[]: This is a quantum register that holds the target qubits.
\item O[]: This is a quantum register that holds the output  qubit.
\end{itemize}
\item \textbf{Quantum part of the algorithm}
\begin{itemize}
\item Quantum portion of the algorithm are executed on quantum computers, where unitary quantum gates (Hadamard, NOT, CNOT) are applied on the input qubits.
 \end{itemize} 
  \end{itemize}


The execution of the algorithm is as follows:
\begin{itemize}
\item \textbf{Step 1}: Initialize Adjacency Matrix for the input graph. Calculate the total number of combinations of vertices using a classical algorithm.
\item \textbf{Step 2}: Initialize quantum register I[] that holds the input qubits with $\ket{0}$.
\item \textbf{Step 3}: Apply Hadamard gates on all the input qubits.
\item \textbf{Step 4}: Apply Hadamard gate on the output qubit.
\item \textbf{Step 5}: Execute the proposed unitary blocks $i.e.,$ Qubit Activation, Clique Detector, Qubit Deactivation for finding the marked states. This checks full connectivity in a set of vertices by applying unitary transformation of quantum gates. 
\item \textbf{Step 6}: Apply Grover's operator for maximizing the amplitude of the marked states.
\item \textbf{Step 7}: Measure the output of quantum register I[] using classical register.

 \end{itemize}

 \subsubsection{Automated and Generalized Algorithm  for Circuit Synthesis of Proposed Oracle for \textit{k}-Clique Problem}
 The proposed algorithm of oracular circuit synthesis for the \textit{k}-clique problem is illustrated in this subsection. Algorithm 1 describes the algorithm for oracle circuit synthesis of the \textit{k}-clique problem. The algorithm takes the adjacency matrix of the given graph as input and the $k$ sized cliques to be searched. The output of the algorithm is the oracle circuit in the form of a circuit netlist.

  \begin{algorithm}[]
  \caption{AutoGenerateOraclek-Clique}
  \scriptsize
\KwIn{Adjacency matrix $adj(n,n)$ of graph $G(V,E)$. Size of the clique to be searched that is \textit{k}. $I_m$ are input qubit lines where $1 \leq m \leq k*\ceil{\log_2 n}$. $A_x$ and $T$ are ancilla lines where $1 \leq x \leq {k \choose 2}$. $O$ is output line.
}

\KwOut{Circuit Netlist(QASM)}
Initialize $I_m$ input lines with $\ket{0}$ followed by Hadamard gate, $A_x$ ancilla lines with $\ket{0}$, $T$ ancilla line with $\ket{0}$, and output line $O$ with $\ket{1}$ followed by a Hadamard gate.
Make $n \choose k$ combinations of vertices and store them in an array $comb_{arr}$. Initialize $comb\leftarrow 0$, $m\leftarrow k*\ceil{\log_2 n}$ and $x\leftarrow 1$\;

 \For{$i\leftarrow 1$ to $n \choose k$}
 {
insert Qubit Activation Block;\\
insert \textit{k}-clique Detector Block;\\
insert Qubit Deactivation Block;\\
$comb \leftarrow comb+k$ (To check next possible combination of vertices for searching $k$-clique),
$x \leftarrow 1$\; 
}

Insert CNOT gate $T$ line as control and $O$ output line as target\;

 For mirror circuit:\\
\For{$i\leftarrow 1$ to $n \choose k$}
{
insert Qubit Activation Block;\\
insert \textit{k}-clique Detector Block;\\
insert Qubit Deactivation Block;\\
$comb \leftarrow comb+k, x \leftarrow 1$\;
}

    \end{algorithm}

\begin{algorithm}[]
\caption{Qubit Activation Block \& Qubit Deactivation Block}
\scriptsize
\label{alg:oneone}

Binary representation of {$comb_{arr}[comb]$, $comb_{arr}[comb+1]$ $\dots$ $comb_{arr}[comb+k-1]$} in $\ceil{\log_2 n}$ binary digits stores in an array $active$, the current possible combination of vertices for searching $k$-clique\;
Qubit Activation Block:\\
\For{$r\leftarrow 0$ to $m-1$}
{
\If{$active(r)\leftarrow 0$}
{
insert NOT gate to qubit line $I_{r+1}$ to make input qubit line active. $r\leftarrow r+1$;
}
\Else{$r\leftarrow r+1$}

}
            
Qubit Deactivation Block:\\  
    \For{$r\leftarrow 0$ to $m-1$}
{
\If{$active(r)\leftarrow 0$}
{
insert NOT gate to qubit line $I_{r+1}$ to make input qubit line active. $r\leftarrow r+1$;
}
\Else{$r\leftarrow r+1$}

}

    \end{algorithm}

\begin{algorithm}[]
\caption{\textit{k}-clique Detector Block}
\scriptsize
\label{alg:onetwo}

    \For{$i\leftarrow comb$ to $comb+k-2$}
    {
        \For{$j\leftarrow i+1$ to $comb+k-1$}
        {
            
            \If{$adj(comb_{arr}(i),comb_{arr}(j))\leftarrow 1$}
            {
            \If{$(j+1)\%k\leftarrow 0$}
            {
                insert a multi-controlled Toffoli gate with $I_{((i\%k)*\ceil{\log_2 n})+1}$ to $I_{((i+1)\%k)*\ceil{\log_2 n}}$ input qubit lines and $I_{((j\%k)*\ceil{\log_2 n})+1}$ to $I_{((j\%k)+1)*\ceil{\log_2 n}}$ input qubit lines as control and ancilla $A_x$ as target. $x\leftarrow x+1$\;
            }
            \Else
            {
             insert a multi-controlled Toffoli gate with $I_{((i\%k)*\ceil{\log_2 n})+1}$ to $I_{((i+1)\%k)*\ceil{\log_2 n}}$ input qubit lines and $I_{((j\%k)*\ceil{\log_2 n})+1}$ to $I_{((j+1)\%k)*\ceil{\log_2 n}}$ input qubit lines as control and ancilla $A_x$ as target. $x\leftarrow x+1$\;
            }
            }
            \Else
            {
        $x\leftarrow x+1$\
            }
        }
    }
    Insert a k-controlled Toffoli gate with each of the k $A_x$ ancilla lines as control and ancilla $T$ as target\;
    $x\leftarrow k$\, $comb\leftarrow 0$\;
 \For{$j\leftarrow comb+k-1$ to $comb+k-2$}
    {
        \For{$i\leftarrow j-1$ to $comb$}
        {
            \If{$adj(comb_{arr}(i),comb_{arr}(j))\leftarrow 1$}
            {
            \If{$(j+1)\%k\leftarrow 0$}
            {
                insert a multi-controlled Toffoli gate with $I_{((i\%k)*\ceil{\log_2 n})+1}$ to $I_{((i+1)\%k)*\ceil{\log_2 n}}$ input qubit lines and $I_{((j\%k)*\ceil{\log_2 n})+1}$ to $I_{((j\%k)+1)*\ceil{\log_2 n}}$ input qubit lines as control and ancilla $A_x$ as target. $x\leftarrow x-1$\;
            }
            \Else
            {
             insert a multi-controlled Toffoli gate with $I_{((i\%k)*\ceil{\log_2 n})+1}$ to $I_{((i+1)\%k)*\ceil{\log_2 n}}$ input qubit lines and $I_{((j\%k)*\ceil{\log_2 n})+1}$ to $I_{((j+1)\%k)*\ceil{\log_2 n}}$ input qubit lines as control and ancilla $A_k$ as target. $x\leftarrow x-1$\;
            }
            }
            \Else{$x\leftarrow x-1$}
        }
    }

\end{algorithm}
 The steps of the algorithms are described as follows: 
 \begin{enumerate}[Step 1:]

\item  From the adjacency matrix, the number of vertices in the graph are obtained, which are further used to calculate the total number of qubit lines required for the oracle circuit. If there are $n$ vertices in the input graph, at first vertices are indexed as $\{0, 1, 2, \dots n-1\}$. The algorithm stores $n \choose k$ combination of vertices without duplicate entry in an array called $comb_{arr}$. The number of qubits required to represent each vertex is $\ceil{\log_2 n}$. Hence, a total of $k*\ceil{\log_2 n}$ qubit lines are required for input. There are two types of ancilla qubit lines, $A_x$ and $T$ which are initialized by $\ket{0}$. $A_x$ ancillas are used for checking whether two vertices are connected by an edge and $T$ ancilla is used for checking if $k \choose 2$ edges form a clique of size \textit{k}. There is one output line $O$, initialized to $\ket{1}$, which indicates if a clique of size \textit{k} exists in the input graph.\\ 
\item The algorithm first applies Hadamard gates on all input qubit lines. For every possible combination from array $comb_{arr}$, binary representation of \textit{k} vertices in $\ceil{\log_2 n}$ binary digits stores in array $active$. This is further required to activate all the input qubit lines having $\ket{0}$ quantum state using appropriate NOT gates to make them suitable as control lines for MCT operation with the help of Qubit Activation method. This Qubit Activation method is described in Algorithm \ref{alg:oneone}.\\

\item For every possible combination from array $comb_{arr}$, the algorithm checks every possible vertex pairs $(comb_{arr}(i),comb_{arr}(j))$ that are connected by an edge using edge detector block and insert a \textit{k}-clique detector block on appropriate qubit lines for respective vertices.
The circuit synthesis of \textit{k}-clique Detector Block is described in Algorithm \ref{alg:onetwo}. The algorithm checks for every possible combination of \textit{k} vertices without duplicate entry, if vertex pairs $(comb_{arr}(i),comb_{arr}(j))$ are connected by an edge, then insert a multi-controlled Toffoli gate with $I_{((i\%k)*\ceil{\log_2 n})+1}$ to $I_{(i\%k)*\ceil{\log_2 n}}$ input qubit lines and $I_{((j\%k)*\ceil{\log_2 n})+1}$ to $I_{((j\%k)+1)*\ceil{\log_2 n}}$ input qubit lines as control and one of the ancilla $A_x$ as target if $(j+1)\%k$ is zero, (where $i, j$ are the index values of array $comb_{arr}$) or else insert a multi-controlled Toffoli gate with $I_{((i\%k)*\ceil{\log_2 n})+1}$ to $I_{(i\%k)*\ceil{\log_2 n}}$ input qubit lines and $I_{((j\%k)*\ceil{\log_2 n})+1}$ to $I_{(j+1)*\ceil{\log_2 n}}$ input qubit lines as control and one of the ancilla $A_x$ as target. If all vertex pairs of a vertex combination are adjacent, then each of the  $A_x$ ancilla qubits are activated, which implies there exists a clique of size k between them. This information is stored in the $T$ ancilla by applying a $k \choose 2$-controlled Toffoli gate with $A_{x}$ ancillas as control and $T$ ancilla as target. These steps are needed to be repeated followed by Qubit Deactivation Block to generate the mirror of it in order to keep the overall circuit reversible. \\

\item The algorithm then applies a CNOT gate to the output line with $T$ ancilla as control. As a result, the output line shall indicate the presence of a clique of size k in the graph. On the contrary, if there is no clique of size k in the input graph, $T$ ancilla will be in their initial state $\ket{0}$.

\par The remaining steps of the algorithm generates the mirror for the original circuit in order to keep the overall circuit netlist reversible. To understand Algorithm 1 more precisely, an example of 4 vertices graph has been considered The oracle circuit generated for the example graph by the proposed algorithm is illustrated in the next subsection.

\end{enumerate}

\subsubsection{Generation of proposed Oracle Circuit for $k$-clique Problem for an Exemplified Graph}
\begin{figure}[ht!]
\centering
\includegraphics[width= 1.5in]{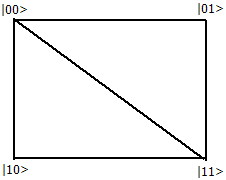}
\caption{Graph with clique size 3}
\label{square}
\end{figure}

\begin{figure}[ht!]
\centering
\includegraphics[width= 5in]{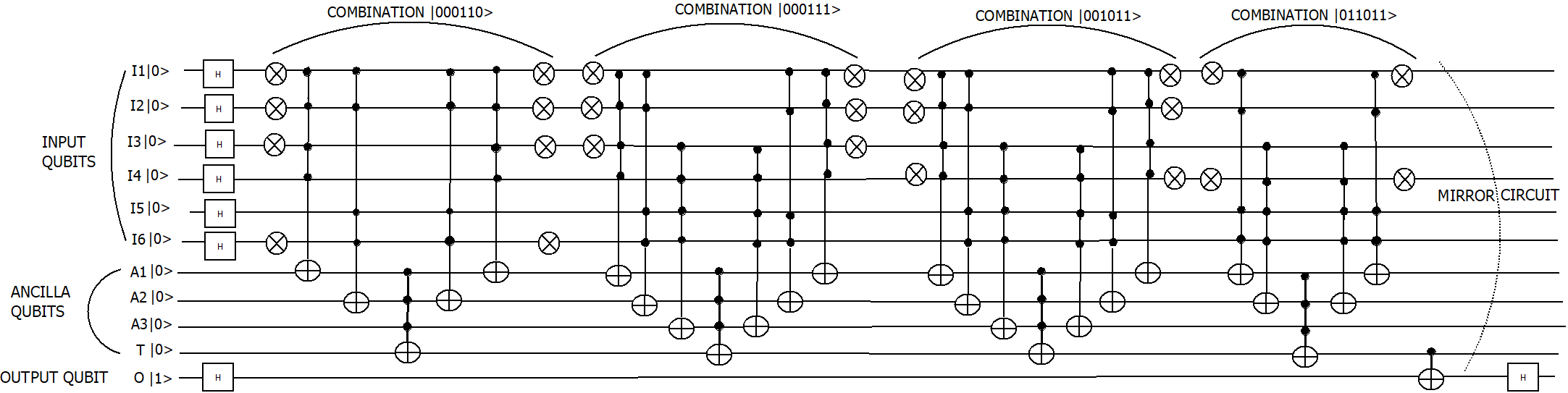}
\caption{Circuit for checking the presence clique of size 3}
\label{clique3}
\end{figure}

The circuit for finding triangle in the graph of Figure \ref{square} has been shown in Figure \ref{clique3}. For a $4$-vertex graph, adjacency matrix is represented by $adj(4, 4)$ and vertices are indexed as ${0, 1, 2, 3}$. Each vertex is represented by $\ceil{\log_2 4}=2$ qubits. Hence, total six qubit lines are required for inputs. As per Algorithm 1, input qubit lines are represented as $I_{m}$ , where $1\leq m \leq 6$. These input qubit lines are initialized with $\ket{0}$ followed by Hadamard gate. In order to store edge information, three ancilla lines $A_{x}$, where $1\leq x \leq 3$, one ancilla line $T$ initialized to $\ket{0}$ and one output line $O$ initialized to $\ket{1}$ followed by Hadamard gate are required to store the output of the circuit. Hence, a total of $11$ qubit lines are required for the simulation. The algorithm now stores $4 \choose 3$ $=$ $4$ possible combination of vertices as $((0, 1, 2), (0, 1, 3), (0, 2, 3), (1, 2, 3))$ in an array called $comb_{arr}$. The algorithm now initializes variable $comb$ as $0$, variable $m$ as $6$ and $x$ as $1$.

\begin{itemize}
\item At first the algorithm considers the first combination of three vertices. Hence, binary representation of three vertices $(0, 1, 2)$ in $\ceil{\log_2 4}=2$ binary digits  as $(000110)$ stores in an array $active$ of length 6.
\item For each zero in $active$ array, insert NOT gate to qubit line $I_{r+1}$ where $r$ is the index value of $active$ array to make these input lines suitable as control for MCT operation or edge detector block.
\item Thereafter, for every possible vertex pair $(0,1)$, $(0,2)$, and $(1,2)$, the algorithm checks that they are connected by an edge or not by accessing the information of adjacency matrix.
\item At first for vertex pair $(comb_{arr}(i=0), comb_{arr}(j=1))=(0,1)$, $adj(0,1)=1$, insert an MCT gate with $I_{1}$ to $I_{2}$ input qubit lines and $I_{3}$ to $I_{4}$ input qubit lines as control and $A_{2}$ as target.

\item In case of vertex pair ($(comb_{arr}(i=0), comb_{arr}(j=2))=(0,2)$, $adj(0,2)=1$, then  $(j+1)\%3$ is zero, then insert an MCT gate with $I_{1}$ to $I_{2}$ input qubit lines and $I_{5}$ to $I_{6}$ input qubit lines as control and $A_{1}$ as target.

\item Similarly, for vertex pair $(comb_{arr}(i=1), comb_{arr}(j=2))=(1,2)$, the entry for the adjacency matrix $adj(1,2)$ is 0 as there is no edge between vertex 1 and 2. So, algorithm does not insert any MCT gate on the $A_{3}$ ancilla line. 
\item Thereafter, Insert a three-controlled Toffoli gate with each of the three $A_{1}$, $A_{2}$, and $A_{3}$ ancilla lines as control and $T$ ancilla as target.
\item Again to make three $A_{1}$, $A_{2}$, and $A_{3}$ ancilla lines reusable, In case of vertex pair $(comb_{arr}(i=0), comb_{arr}(j=1))=(0,1)$, $adj(0,1)=1$, insert an MCT gate with $I_{1}$ to $I_{2}$ input qubit lines and $I_{3}$ to $I_{4}$ input qubit lines as control and $A_{2}$ as target.
\item for vertex pair ($(comb_{arr}(i=0), comb_{arr}(j=2))=(0,2)$, $adj(0,2)=1$, then  $(j+1)\%3$ is zero, then insert an MCT gate with $I_{1}$ to $I_{2}$ input qubit lines and $I_{5}$ to $I_{6}$ input qubit lines as control and $A_{1}$ as target.

\item These steps are repeated for other three combination of three vertices. The algorithm then applies a CNOT gate to the output line $O$ with $T$ ancilla as control. As a result, the output line shall indicate the presence of a clique in the graph. The remaining steps of the algorithm generate the mirror for the original circuit in order to keep the overall circuit netlist reversible.
\end{itemize}

\subsubsection{Mathematical Formulation of Proposed Oracle for \textit{k}-clique Problem for an Exemplified Graph}
Generalization of the oracle for $k$-clique problem is already well established in the previous subsections. This section shows how an arbitrary graph formulates the proposed oracle mathematically.
We have considered the same graph as shown in Figure \ref{square} and the corresponding generated circuit given in Figure \ref{clique3}.

\textbf{Initialization:}
The input qubits are initialized with $\ket{0}$. Ancilla qubits are initialized with $\ket{0}$. Output qubit is initialized with $\ket{1}$. Entire initialization can be mathematically written as:
\[\vert\psi\rangle\otimes\vert\theta\rangle\otimes\vert\phi\rangle=\vert 0\rangle^{\otimes 6}\otimes\vert 0\rangle^{\otimes 4}\otimes\vert 1\rangle\]

 \textbf{Hadamard Transformation:}
After the initialization, the Hadamard transform $H^{\otimes 6}$ on input qubits and $H$ on output qubit is performed, therefore all possible states are superposed as $\ket{\psi_{0}}\otimes\ket{\theta_{0}}\otimes\ket{\phi_{0}}$, where

\[\vert\psi_{0}\rangle=\,{1\over\sqrt{2^{6}}}\sum_{i=0}^{2^{6}-1}\vert i\rangle\]

\[\ket{\theta_{0}}=\ket{0000}\]

\[\vert\phi_{0}\rangle=\,{1\over\sqrt{2}}\left(\vert 0\rangle-\vert 1\rangle\right)\]

\[\ket{\rho_0}=\vert\psi_{0}\rangle\otimes\vert\theta_{0}\rangle\otimes\vert\phi_{0}\rangle={1\over\sqrt{2^{6}}}\sum_{i=0}^{2^{6}-1}\vert i\rangle\otimes\vert 0\rangle^{\otimes 4}\otimes{1\over\sqrt{2}}\left(\vert 0\rangle-\vert 1\rangle\right)\]

\[\implies {\frac{1}{\sqrt{2^{6}}}}(\ket{000000}+ \dots + \ket{111111})\otimes\vert 0\rangle^{\otimes 4}\otimes{1\over\sqrt{2}}\left(\vert 0\rangle-\vert 1\rangle\right) \]

\[\implies {1\over\sqrt{2^{7}}}(\ket{00000000000}+\ket{00000100000}+ \dots + \ket{11111100000})\] \[+{1\over\sqrt{2^{7}}}(-\ket{00000000001}-\ket{00000100001}- \dots - \ket{11111100001})\]

\textbf{Oracle:}
\begin{itemize}
\item \textbf{1st combination of vertices $\ket{000110}$}:- 
\begin{enumerate}

\item \textbf{Search space reduction using qubit activation}:- After applying NOT gates to the 1st ($I_1$), 2nd ($I_2$), 3rd ($I_3$) and 6th ($I_6$) qubit of all superposed states, the quantum state evolves as \\ \\
$\ket{\rho_1}= {\frac{1}{\sqrt{2^{7}}}} (\ket{11100100000}+\ket{11100000000}+ \ket{11101100000} +\ket{11101000000}+ \ket{11110100000} + \ket{11110000000}+\ket{11110000000} +\mathbf{\ket{11111100000}}+\dots+\ket{00011000000})+{1\over\sqrt{2^{7}}}(-\ket{11100100001}-\ket{11100000001}- \dots\mathbf{\ket{11111100001}}\dots - \ket{00011000001})$ \\

\item \textbf{Edge detection}:- Two MCT gates are applied for the combination of vertices ($\ket{00}$, $\ket{10}$) and ($\ket{01}$, $\ket{10}$) for two edges. The quantum state evolves as  \\ \\ $\ket{\rho_2}= {\frac{1}{\sqrt{2^{7}}}} (\ket{11100100000}+\ket{11100000000}+ \ket{11101100000} +\ket{11101000000}+ \ket{11110100000} + \ket{11110000000}+\ket{11110000000} +\mathbf{\ket{11111111000}}+\dots+\ket{11111100000})+{1\over\sqrt{2^{7}}}(-\ket{11100100001}-\ket{11100000001}- \dots\mathbf{\ket{11111111001}}\dots - \ket{00011000001})$ \\

\item \textbf{Clique detection}:- An MCT gate flips the fourth ancilla qubit $T$, if the three ancilla qubits $A_1$, $A_2$, $A_3$ are 1. The quantum state evolves as \\ \\ 
$\ket{\rho_3}= {\frac{1}{\sqrt{2^{7}}}} (\ket{11100100000}+\ket{11100000000}+ \ket{11101100000} +\ket{11101000000}+ \ket{11110100000} + \ket{11110000000}+\ket{11110000000} +\mathbf{\ket{11111111000}}+\dots+\ket{11111100000})+{1\over\sqrt{2^{7}}}(-\ket{11100100001}-\ket{11100000001}- \dots\mathbf{\ket{11111111001}}\dots - \ket{00011000001})$ \\
\item\textbf{Mirror circuit for 1st combination}: After applying mirror circuit the quantum state evolves as\\ \\
 $\ket{\rho_4}={\frac{1}{\sqrt{2^{7}}}} (\ket{00000000000}+\ket{00000100000}+ \ket{00001000000} +\ket{00001100000}
\ket{00010000000} + \ket{00010100000}+\ket{00010100000}+\mathbf{\ket{00011000000}} +\ket{00011100000}+\dots+\ket{11111100000})+{1\over\sqrt{2^{7}}}(-\ket{00000000001}-\ket{00000100001}- \dots\mathbf{\ket{00011000001}}\dots - \ket{11111100001})$ \\
\end{enumerate}
\item \textbf{2nd combination of vertices $\ket{001011}$}. \\
\begin{enumerate}

\item \textbf{Search space reduction using qubit activation}:- After applying NOT gates to the 1st ($I_1$), 2nd ($I_2$) and 4th ($I_4$) qubit of all superposed states, the quantum state evolves as\\ \\
$\ket{\rho_5}={\frac{1}{\sqrt{2^{7}}}} (\ket{11010000000}+\ket{11010100000}+ \ket{11011000000} +\ket{11001100000} + \dots+ \mathbf{\ket{11000000000}}+ \dots+ \mathbf{\ket{11111100000}} \dots + \ket{00101100000}) + {1\over\sqrt{2^{7}}} (-\ket{11010000001}- \ket{11010100001}- \dots -\mathbf{\ket{11001000001}} \dots -\mathbf{\ket{11111100001}} \dots - \ket{00101100001})$ \\

\item \textbf{Edge Detection}:- Three MCT gates are applied for the combination of vertices ($\ket{00}$, $\ket{10}$), ($\ket{01}$, $\ket{10}$) and ($\ket{00}$, $\ket{01}$) for three edges. The quantum state evolves as\\ \\
 $\ket{\rho_6}={\frac{1}{\sqrt{2^{7}}}} (\ket{11010000000}+\ket{11010100000}+ \ket{11011000000} +\ket{11001100000} +\dots+\mathbf{\ket{11001000000}}+\dots\mathbf{\ket{11111111100}} \dots+\ket{00101100000})+{1\over\sqrt{2^{7}}}(-\ket{11010000001}-\ket{11010100001}- \dots -\mathbf{\ket{11001000001}}- \dots -\mathbf{\ket{11111111101}} \dots - \ket{00101100001})$  \\
\item \textbf{Clique Detection}:- An MCT gate flips the fourth ancilla qubit $T$, if the three ancilla qubits $A_1$, $A_2$, $A_3$ are 1. The quantum state evolves as\\ \\
 $\ket{\rho_7}={\frac{1}{\sqrt{2^{7}}}} (\ket{11010000000}+\ket{11010100000}+ \ket{11011000000} +\ket{11001100000} +\dots+\mathbf{\ket{11001000000}}+\dots\mathbf{\ket{11111111110}} \dots+\ket{0010110000})+{1\over\sqrt{2^{7}}}(-\ket{11010000001} -\ket{11010100001} \dots  -\mathbf{\ket{11001000001}} - \mathbf{\ket{11111111111}} \dots - \ket{00101100001})$ \\ 
\item \textbf{Mirror circuit of 2nd combination}: After applying mirror circuit the quantum state evolves as \\ \\
 $\ket{\rho_8}={\frac{1}{\sqrt{2^{7}}}} (\ket{00000000000}+\ket{00000100000}+ \ket{00001000000}+\dots+\mathbf{\ket{00011000000}}+\dots+\mathbf{\ket{00101100010}}+\dots+\ket{11111100000})+{1\over\sqrt{2^{7}}}(-\ket{00000000001}-\ket{00000100001}-\dots-\mathbf{\ket{00011000001}}-\dots-\mathbf{\ket{00101100011}}-\dots-\ket{11111100001})$  \\
\end{enumerate}
\item \textbf{3rd combination of vertices $\ket{011011}$}:\\
\begin{enumerate}
\item \textbf{Search space reduction using qubit activation}:- After applying NOT gates to the 1st ($I_1$) and 4th ($I_4$) qubit of all superposed states, the quantum state evolves as \\ \\ $\ket{\rho_9}= {\frac{1}{\sqrt{2^{7}}}} (\ket{10010000000}+\ket{10010100000}+ \ket{10011000000}\dots\mathbf{\ket{10001000000}}\dots+\mathbf{\ket{10111100010}} \dots +\mathbf{\ket{11111100000}}+ \dots +\ket{01101100000}) +{1\over\sqrt{2^{7}}}(-\ket{10010000001}-\ket{10010100001}- \dots\mathbf{\ket{10001000001}}-\mathbf{\ket{10111100011}}\dots-\mathbf{\ket{11111100001}}\dots - \ket{01101100001})$ \\

\item \textbf{Edge Detection}:- Two MCT gates are applied for the combination of vertices ($\ket{01}$, $\ket{11}$) and ($\ket{10}$, $\ket{11}$) for two edges. The quantum state evolves as \\ \\
 $\ket{\rho_{10}}={\frac{1}{\sqrt{2^{7}}}} (\ket{10010000000}+\ket{10010100000}+ \ket{10011000000}+\dots+\mathbf{\ket{10001000000}}+\dots+\mathbf{\ket{10111100010}}+\dots+\mathbf{\ket{11111111000}}+\dots+\ket{01101100000}) +{1\over\sqrt{2^{7}}}(-\ket{10010000001}-\ket{10010100001}- \dots-\mathbf{\ket{10001000001}}-\mathbf{\ket{10111100011}}-\dots-\mathbf{\ket{11111111001}}-\dots-\ket{01101100001})$ \\
 
 \item \textbf{Clique Detection}:- An MCT gate flips the fourth ancilla qubit $T$, if the three ancilla qubits $A_1$, $A_2$, $A_3$ are 1. The quantum state evolves as \\ \\
 $\ket{\rho_{11}}={\frac{1}{\sqrt{2^{7}}}} (\ket{10010000000}+\ket{10010100000}+ \ket{10011000000}+\dots+\mathbf{\ket{10001000000}}+\dots+\mathbf{\ket{10111100010}}+\dots+\mathbf{\ket{11111111000}}+\dots+\ket{01101100000}) +{1\over\sqrt{2^{7}}}(-\ket{10010000001}-\ket{10010100001}-\dots-\mathbf{\ket{10001000001}}-\mathbf{\ket{10111100011}}-\dots-\mathbf{\ket{11111111001}}-\dots - \ket{01101100001})$ \\
\item \textbf{Mirror circuit of 3rd combination} After applying mirror circuit the quantum state evolves as\\ \\
$\ket{\rho_{12}}={\frac{1}{\sqrt{2^{7}}}} (\ket{00000000000}+\ket{00000100000}+ \dots+\mathbf{\ket{00011000000}}+\dots +\mathbf{\ket{00101100010}} \dots+\mathbf{\ket{01101100000}}+\dots+\ket{11111100000})+{1\over\sqrt{2^{7}}}(-\ket{00000000001}-\ket{00000100001}- \dots\mathbf{\ket{00011000001}}-\mathbf{\ket{00101100011}}\dots-\mathbf{\ket{01101100001}}\dots - \ket{11111100001})$\\

\end{enumerate}
\item \textbf{4th combination for the input qubits $\ket{000111}$} \\
\begin{enumerate}

\item \textbf{Search space reduction using qubit activation}:- After applying NOT gates to the 1st ($I_1$), 2nd ($I_2$) and 3rd ($I_3$) qubits of all superposed states. The quantum state evolves as \\ \\
$\ket{\rho_{13}}={\frac{1}{\sqrt{2^{7}}}} (\ket{11100000000}+\ket{11100100000}+ \dots+\mathbf{\ket{11111000000}}+\dots\mathbf{\ket{11111100000}}\dots +\mathbf{\ket{11001100010}} \dots+\mathbf{\ket{10001100000}}+\dots+\ket{11111100000})+{1\over\sqrt{2^{7}}}(-\ket{00000000001}-\ket{00000100001}- \dots\mathbf{\ket{00011000001}}-\dots\mathbf{\ket{11111100000}}\dots\mathbf{\ket{00101100011}}\dots\mathbf{\ket{01101100001}}\dots - \ket{11111100001})$ \\

\item \textbf{Edge Detection}:- Three MCT Gates are applied for the combination of the vertices ($\ket{00}, \ket{01}$), ($\ket{00}, \ket{11}$) and ($\ket{01}, \ket{11}$) for three edges. The quantum state evolves as\\ \\
$\ket{\rho_{14}}={\frac{1}{\sqrt{2^{7}}}} (\ket{11100000000}+\ket{11100100000}+ \dots+\mathbf{\ket{11111000000}}+\dots\mathbf{\ket{11111111100}}\dots +\mathbf{\ket{11001100010}} \dots+\mathbf{\ket{10001100000}}+\dots+\ket{11111100000})+{1\over\sqrt{2^{7}}}(-\ket{00000000001}-\ket{00000100001}- \dots\mathbf{\ket{00011000001}}-\dots-\mathbf{\ket{11111111100}}\dots\mathbf{\ket{00101100011}}\dots\mathbf{\ket{01101100001}}\dots - \ket{11111100001})$ \\
\item \textbf{Clique Detection}:- An MCT gate flips the fourth ancilla qubit $T$, if the three ancilla qubits $A_1$, $A_2$, $A_3$ are 1. The quantum state evolves as \\ \\
$\ket{\rho_{15}}={\frac{1}{\sqrt{2^{7}}}} (\ket{11100000000}+\ket{11100100000}+ \dots+\mathbf{\ket{11111000000}}+\dots\mathbf{\ket{11111111110}}\dots +\mathbf{\ket{11001100010}} \dots+\mathbf{\ket{10001100000}}+\dots+\ket{11111100000})+{1\over\sqrt{2^{7}}}(-\ket{00000000001}-\ket{00000100001}- \dots\mathbf{\ket{00011000001}}-\dots-\mathbf{\ket{11111111111}}\dots\mathbf{\ket{00101100011}}\dots\mathbf{\ket{01101100001}}\dots - \ket{11111100001})$ \\
\item \textbf{Mirror circuit of 4th combination}:- After applying mirror circuit the quantum state evolves as\\ \\
$\ket{\rho_{16}}={\frac{1}{\sqrt{2^{7}}}} (\ket{00000000000}+\ket{00000100000}+ \dots+\mathbf{\ket{00011000000}}+\dots +\mathbf{\ket{00101100010}} \dots+\mathbf{\ket{00011100010}}\dots+\mathbf{\ket{01101100000}}+\dots+\ket{11111100000})+{1\over\sqrt{2^{7}}}(-\ket{00000000001}-\ket{00000100001}- \dots\mathbf{\ket{00011000001}}-\dots-\mathbf{\ket{00011100011}}\dots\mathbf{\ket{00101100011}}\dots\mathbf{\ket{01101100001}}\dots - \ket{11111100001})$ \\

\end{enumerate}

\item \textbf{Flip the output qubit:}

CNOT gate is applied on the output qubit ($O$) with the fourth ancilla qubit ($T$) as control. \\ \\
$ \ket{\rho_{17}}={\frac{1}{\sqrt{2^{7}}}} (\ket{00000000000}+\ket{00000100000}+ \dots+\mathbf{\ket{00011000000}}+\dots +\mathbf{\ket{00101100011}} \dots+\mathbf{\ket{00011100011}}\dots+\mathbf{\ket{01101100000}}+\dots+\ket{11111100000})+{1\over\sqrt{2^{7}}}(-\ket{00000000001}-\ket{00000100001}- \dots\mathbf{\ket{00011000001}}-\dots-\mathbf{\ket{00011100010}}\dots\mathbf{\ket{00101100010}}\dots\mathbf{\ket{01101100001}}\dots - \ket{11111100001})$\\

\item \textbf{Hadamard gate on the output qubit:} After applying Hadamard gate on the output qubit ($O$) the quantum state evolves as\\ \\
$\ket{\rho_{18}}= {\frac{1}{\sqrt{2^{7}}}} (\ket{00000000001}+\ket{00000100001}+ \dots+\ket{00011000001} \dots+\ket{01101100001}\dots+\ket{11111100001})+{1\over\sqrt{2^{7}}}(-\mathbf{\ket{00011100011}}-\mathbf{\ket{00101100011}})$ \\
\par We observe that the two marked states $\ket{000111}$ and $\ket{001011}$ are with negative amplitude only. Now, Grover's diffusion operator performs the inversion about the average of all amplitudes of the quantum state for a certain number of iterations to get the amplitude of the marked state large enough. So, the amplitude of the marked states become higher than other states.
\end{itemize}
\subsection{Automated and Generalized Algorithm for Solving Maximum Clique Problem using the Proposed Approach of \textit{k}-clique Problem in Classical-Quantum Hybrid Architecture}
\begin{figure}[ht!]
\centering
\includegraphics[width=5.5in]{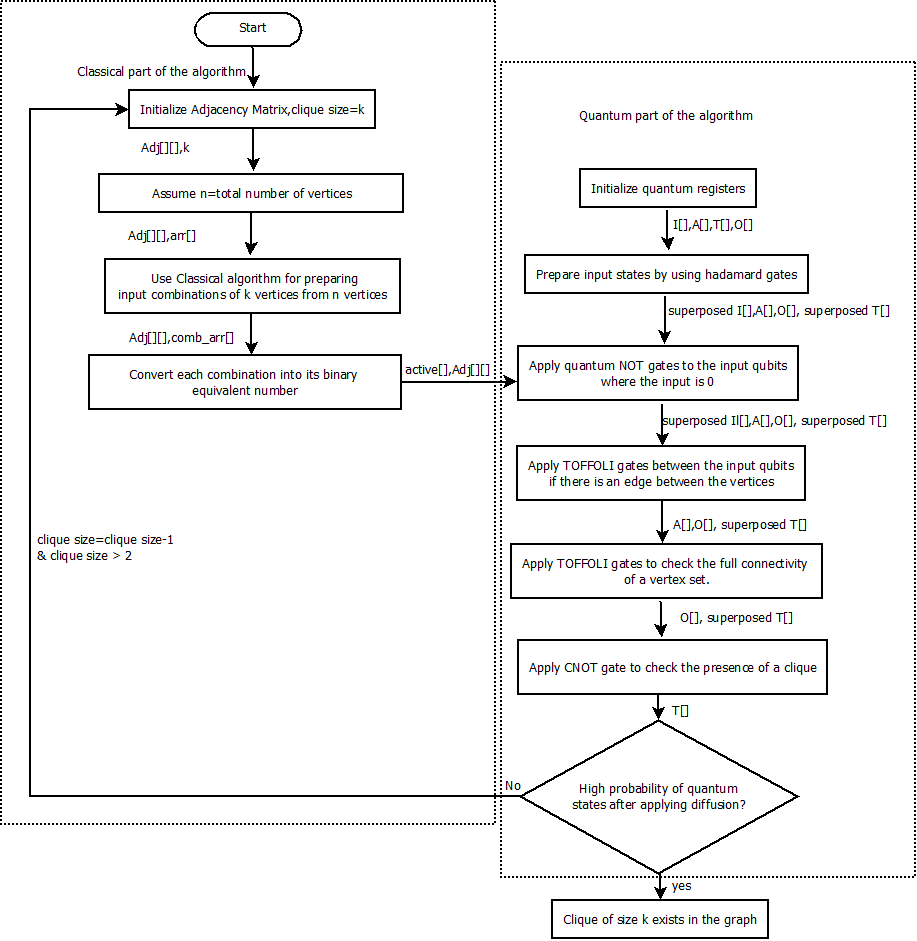}
\caption{The flow of the complete algorithm for maximum clique problem in classical-quantum hybrid architecture}
\label{maxcliquealgo}
\end{figure}

The automated and generalized algorithm of quantum circuit synthesis for maximum clique problem is discussed in this subsection. Maximum clique problem finds clique in a graph with maximum cardinality. Input of the algorithm is the Adjacency Matrix of the graph $G(V,E)$ and the output is the QASM gatelist. The algorithm uses the concept of proposed \textit{k}-clique algorithm. The algorithm starts with finding clique of size $k=n$ ($n$ is total number of vertices of the graph) using the algorithm of \textit{k}-clique. If a clique is found, it is maximum clique of the graph. Otherwise a clique is searched for size $n-1$. The process continues up to $n=2$. The algorithm is performed in classical-quantum hybrid architecture. As described in subsection 2.4, hybrid architecture requires a feedback loop from quantum part of the algorithm to the classical part of the algorithm, which is portrayed in Figure \ref{maxcliquealgo} for MCP problem.

The steps of the circuit synthesis for MCP are shown below:
  
\begin{itemize}
  \item \textbf{Step 1}: Algorithm for finding \textit{k}-clique is used for searching clique of size $n$, where the number vertices of the given graph is $k$.

  \item \textbf{Step 2}: If a clique is found, then exit from the process. 
  \item \textbf{Step 3}: If clique is not found, then clique size $n$ is reduced to $n-1$ and repeat the whole process from step 1.
  \item \textbf{Step 4}: At each iteration the value of $n$ is reduced by 1 $(n, n-1,n-2,n-3,\dots, 2)$, until a clique is found.

\end{itemize} 

 The algorithm for solving MCP using the proposed approach of $k$-clique problem is illustrated in Algorithm 4.

  \begin{algorithm}[]
  
  \caption{AutoGenerateOracleMaximumClique}
  \scriptsize
  \KwIn{Adjacency Matrix Adj[][]} 
  	\KwOut{A file ”OutputMaximumClique.QASM”}
  	 keep all vertices of the input graph($n$) in an array $arr[]$ \\
  	  	\For {$i \leftarrow n$ downto $2$}  
  	  		{  	  		
  	  		\textbf{Use \textit{k}-clique alogorithm for finding clique of size $i$} \\
  	  		 \If {clique of size $i$ is found}
  	  		 {
			 		\textbf{Maximum clique is found of size $i$ } \\
			         \textbf{Exit from the loop} \\
			}
  	  		}
  	
\end{algorithm} 
\

\section{Cost analysis of the Circuit}
Cost of a quantum circuit depends on the number of the input lines, ancilla lines and the number of gates used. If the circuit is designed for a size $k$, then total combination is $n \choose k$, when the number of vertices of given graph is $n$. Therefore, a total number of $O(k * {n \choose k} +k)$ logical gates are required to design the oracle circuit for a variant of $k$-clique problem, which lists all the cliques of size $k$. As this is a first its kind approach of solving such $k$-clique problem, there are no sufficient works to compare the oracular gate cost. Albeit we assume that if the oracle circuit of the state-of-the-art work \cite{sota} can be designed in such a way that it gives all the cliques of size $k$ as output, then the oracular gate cost of our proposed work should be better than \cite{sota}. But, this proof remains out of the scope of this paper. In present scenario, due to the constraint of less number of qubits, quantum circuit with large number of gates with multiple qubits cannot be implemented in real hardware due to its depth. As discussed in the previous section that each vertex is represented by $\ceil{log_2n}$ qubits, so total number of input qubits $k*\ceil{log_2n}$ are required along with ${k \choose 2}+2$ ancilla qubits as shown in Table \ref{circuitcost}. Therefore, for detecting an edge between two vertices, we need a multi-controlled Toffoli gate with $2*\ceil{log_2n}$ control qubits and for  detection of clique of size \textit{k}, we need multi-controlled Toffoli gate with $k \choose 2$ control qubits as shown in Table \ref{circuitcost}. Finally, we can conclude that $k*\ceil{log_2n}$-qubit MCT gate is required to perform diffusion part of the proposed approach of Grover's algorithm.
\begin{table*}[ht!]
\caption{Quantum Cost Analysis of Oracular Circuit for $k$-clique Problem}
\centering
\begin{tabular}{| p{3.5em} | p{5.5em} | p{4.5em} | p{13.5em} |}
 \hline
 No of vertices & No of input qubit & No of ancilla qubits & Maximum number of controls in the MCT Gate \\
 \hline
   n & $k*\ceil{log_2n}$ & $k \choose 2$ + 2 &  \shortstack{Edge detection=$2*\ceil{log_2n}$,\\ Clique detection=$k \choose 2$}\\
  \hline
\end{tabular}
\label{circuitcost}
\end{table*}

Analysis of the qubit cost and the depth of the complete circuit for $k$-clique problem is as follows:

\begin{lemma}
Improved qubit cost for proposed approach of solving $k$-clique problem, when $n >> k$, where $n$ is a large number of nodes of given graph ($G(V=n, E=e)$) and $k$ is respectively very small.
\end{lemma}
The total number of qubits are required to solve $k$-clique problem using proposed approach is $k*\ceil{log_2n}$ $+$ $k \choose 2$ $+$ $2$. Whereas discussed in subsection 2.6.2 in the background, state-of-the-art approach requires a total $n$ $+$ $\ceil{log_2 k}$ $+$ $3$ qubits for solving $k$-clique problem, which gives a clique of size $k$ as output. The state-of-the-art problem is the subset of the proposed problem of $k$-clique as our proposed algorithm  gives all the cliques of size $k$ as output. So from this, we can infer that if we solve such variant of $k$-clique problem, then it can also solves the state-of-the-art problem of $k$-clique. To prove the stated $Lemma$, we have considered three instances of $n$ and $k$ values. Let's start with if $n=1024$ and respectively $k$ is very small $i.e.,$ 3, then the qubit cost $3*\ceil{log_2 1024}$ $+$ $3 \choose 2$ $+$ $2$ $=$ $35$ of the proposed approach is very efficient as compared to the qubit cost $1024$ $+$ $\ceil{log_2 3}$ $+$ $3$ $=$ $1029$ of state-of-the-art approach. Now, if we slightly increase the value of $k$ as 32, but still remains very small as compared to $n=1024$, then the qubit cost $32*\ceil{log_2 1024}$ $+$ $32 \choose 2$ $+$ $2$ $=$ $818$ of the proposed approach is still efficient as compared to the qubit cost $1024$ $+$ $\ceil{log_2 32}$ $+$ $3$ $=$ $1032$ of state-of-the-art approach. But, if we increase the value of $k$ little further towards $n$, our approach becomes costlier with respect to qubit cost to find a clique size of $k$ in a given graph. For instance, suppose $n=1024$ and $k=64$, the qubit cost $64*\ceil{log_2 1024}$ $+$ $64 \choose 2$ $+$ $2$ $=$ $2658$ of the proposed approach is costlier as compared to the qubit cost $1024$ $+$ $\ceil{log_2 64}$ $+$ $3$ $=$ $1033$ of state-of-the-art approach. Thus, we can conclude that our proposed approach outperforms the state-of-the-art approach of solving $k$-clique problem with respect to qubit cost, when $k$ is respectively very small with respect to large $n$.

\begin{proposition}
Improved circuit depth for proposed approach of solving $k$-clique problem, when $n >> k$, where $n$ is a large number of nodes of given graph ($G(V=n, E=e)$) and $k$ is respectively very small.
\end{proposition}

From $Lemma ~ 1$, it can be inferred that the input qubit cost $k*\ceil{log_2n}$ of the proposed approach is also efficient as compared to the input qubit cost $n$ of the state-of-the-art approach, when $n >> k$ and when $k \equiv n$, it becomes inefficient as $n*\ceil{log_2n}$ $>$ $n$. It is already defined that the diffusion operator is imposed on all the input qubits. Therefore, a $k*\ceil{log_2n}$-qubit MCT is required to perform this diffusion for the proposed approach as compared to $n$-qubit MCT for the state-of-the-art. As $k*\ceil{log_2n}$ $<$ $n$, when $n >> k$, we can infer from Figure \ref{dvtc} in subsection 2.3 that the depth of the circuit is also efficient of the proposed approach as compared to the state-of-the-art method. As Figure \ref{dvtc} suggests that with the increasing number of controls in the MCT gate also increases the depth of the circuit drastically, our goal is to restrict the number of controls in the MCT gate as compared to the \cite{sota}, which is successfully achieved for $n >> k$.

\section{Implementation Results}
Quantum simulators are used to simulate the behaviour of a quantum algorithm on a classical computer in order to have an idea of how the circuit is executed on a real quantum computer could be. We can also experiment quantum algorithm in quantum computers via IBM cloud. IBM offers its Quantum Experience platform with freely accessible quantum computers
up to 14 qubits and a quantum simulator up to 32 qubits. Though, we can perform the experimentation with real quantum computers, simulators still play an important role in testing quantum algorithms. We have to consider the fact that, it is not possible to access to certain data which is useful for testing purposes on a real quantum computer. Another problem is that we would like to test a quantum algorithm that needs a large number of qubits, which is still unreachable for real hardware machines. Implementation of a circuit in real quantum device depends on the circuit depth or length of the critical path of the circuit. Unfortunately, the circuit depth is highly dependent
on the hardware layout of qubits and the connections between them. Due to this limitation, we could not implement the circuit for larger graphs in the real quantum device.

\subsection{Ideal Simulation on $ibmq\_qasm\_simulator$}
\begin{figure}[ht!]
\centering
\includegraphics[width= 1.5in]{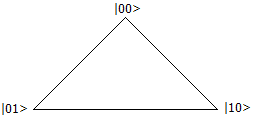}
\caption{Graph with clique size 3}
\label{triangle}
\end{figure}

\begin{figure}[ht!]
\centering
\includegraphics[width= 5in]{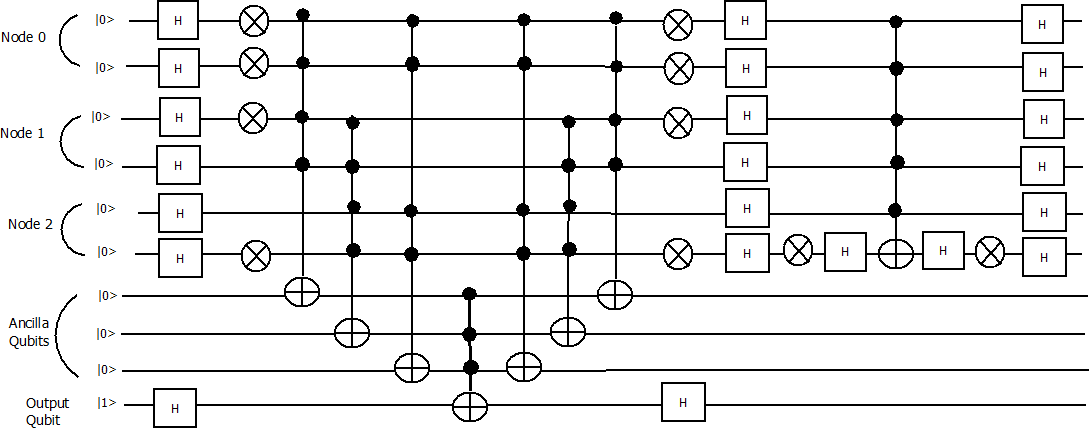}
\caption{circuit for checking the presence clique of size 3}
\label{cliquetriangle}
\end{figure}

The oracle circuit for $k$-clique problem for the graph shown Figure in \ref{triangle} has been implemented in $ibmq\_qasm\_simulator$.
Here, we have 6 qubit lines for inputs, there can be a total of $2^6=64$ different qubit combinations as $\ket{000000}$, $\ket{000001}$, $\dots$, $\ket{111111}$ after applying Hadamard gate to all input qubits. Hence, the database on which Grover's algorithm is applied contains 64 different elements. Figure \ref{cliquetriangle} shows the complete gate level synthesis of the Grover's circuit for 3-clique problem. 

\begin{figure}[ht!]
\centering
\includegraphics[width= 5in]{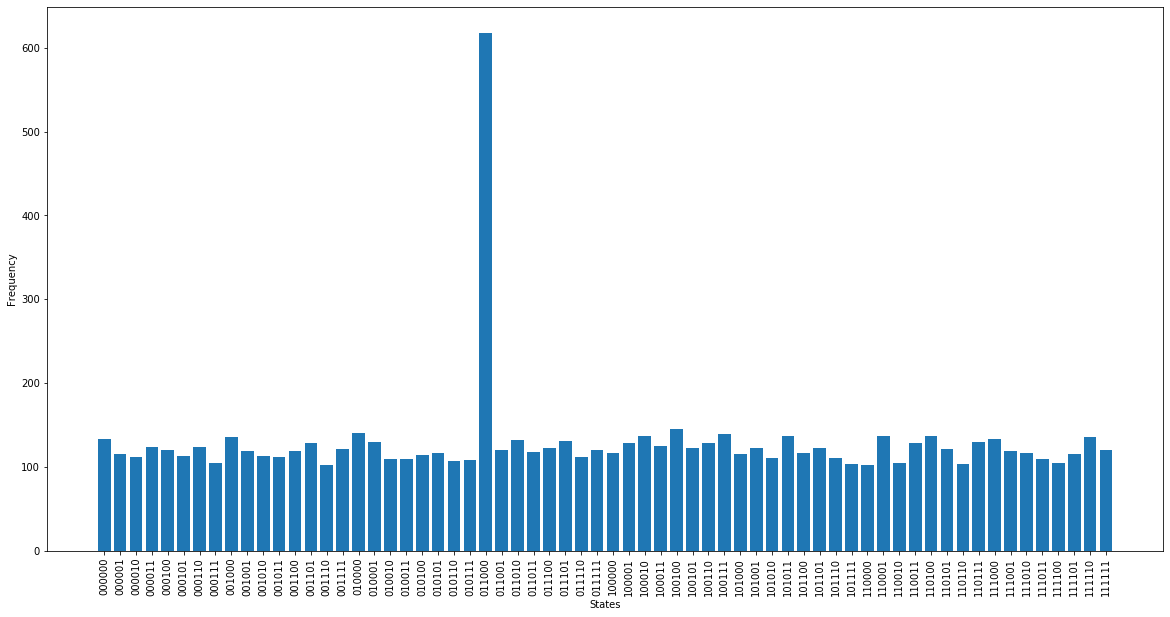}
\caption{Ideal simulation of the algorithm}
\label{outsimu}
\end{figure}

The simulation steps on $IBMQ\_Qasm\_Simulator$ \cite{28} required for the synthesis are as follows:
 \begin{enumerate}
\item The oracle checks whether two vertices are adjacent or not. It determines whether a clique is formed by a three-vertex combination. three-vertex combinations are $\ket{000110}$. It then inverts the amplitude of the solution state for which the vertex combination forms a clique. The solution state for this case is $\ket{000110}$.
\item The output of the oracle is acted upon by the Grover's diffusion operator as shown in Figure \ref{cliquetriangle}. The diffusion operator amplifies the amplitude of the marked solution state.
\item Steps 1 and 2 constitute the Grover's operator for the Grover's search algorithm. Hence, these two steps are repeated $\dfrac{\pi}{4}\sqrt{1}$ times. (For an $N$ item database with $M$ solutions, Grover's iteration must be repeated $\dfrac{\pi}{4}\sqrt{\dfrac{N}{M}}$ times in order to obtain a solution.
\end{enumerate}
The resultant output after applying Grover's operator $\dfrac{\pi}{4}\sqrt{1}$ $\sim$ $O(1)$ time is shown in Figure \ref{outsimu}, where the amplitude of the solution state has been amplified. The location of the solution state is $\ket{000110}$, which is the vertex combination in an example graph of Figure \ref{triangle} that forms clique with high probability. The circuit for clique size three has been simulated in $IBMQ\_Qasm\_Simulator$. The marked state $\ket{011000}$ is in high amplitude as shown in the output Figure \ref{outsimu}, which depicts that $\ket{00}$, $\ket{01}$ and $\ket{10}$ form a triangle.

\subsection{Implementation on $IBMQ\_16\_melbourne$}
\begin{figure}[ht!]
\centering
\includegraphics[width= 5in]{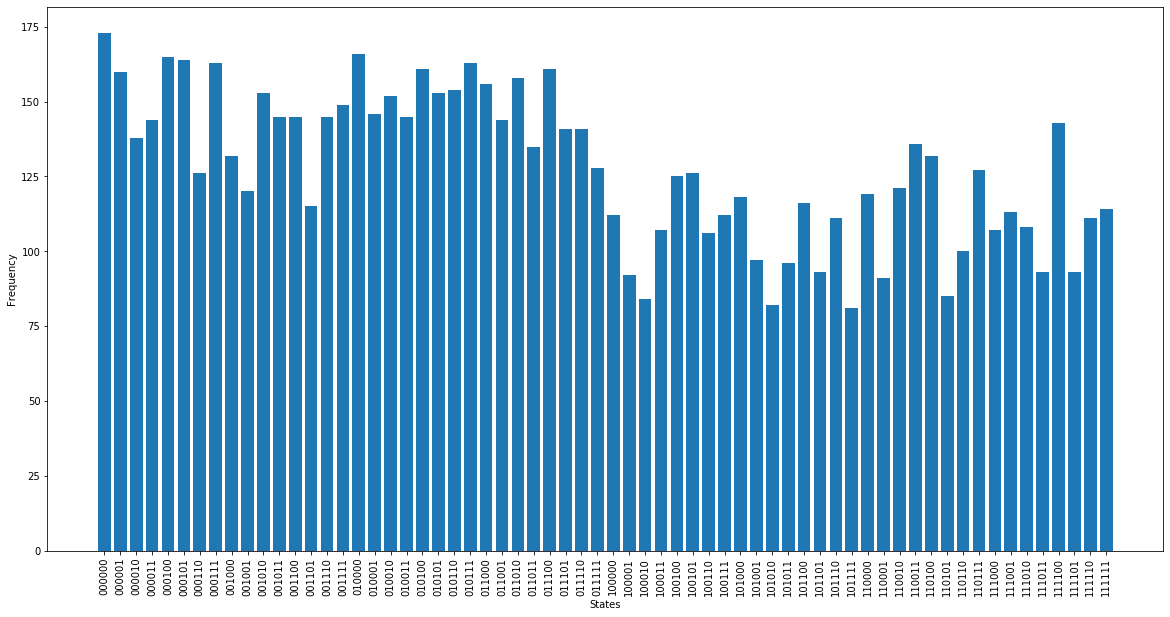}
\caption{Simulation of the algorithm in $IBMQ\_16\_melbourne$}
\label{outmel}
\end{figure}

The results of the simulation using the IBM's simulator are not realistic as it show ideal output without any error and noise. The circuit of Figure \ref{cliquetriangle} is implemented again on $IBMQ\_16\_melbourne$. Executing again the algorithm in $IBMQ\_16\_melbourne$ gives us the histogram in Figure \ref{outmel}. As we observe that the desired state $\ket{000110}$ is not marked properly and the noise caused a loss of information.
\subsection{Simulation using Noise Model of $IBMQ\_16\_melbourne$}
\begin{figure}[ht!]
\centering
\includegraphics[width= 5in]{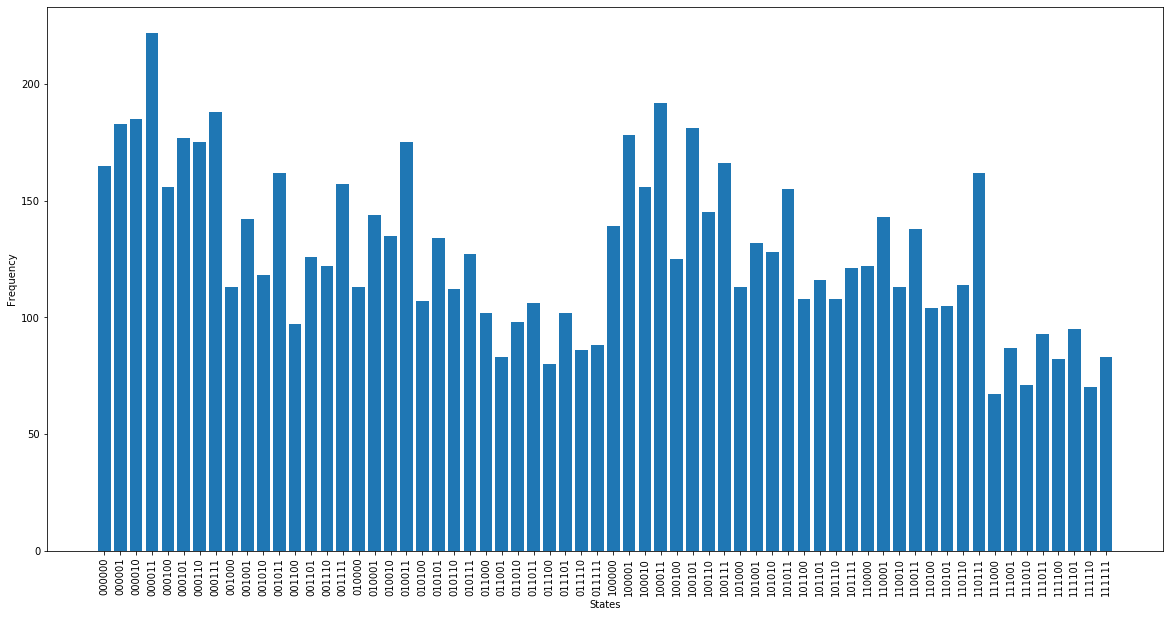}
\caption{Simulation of the algorithm with noise model}
\label{outnoise}
\end{figure}

\begin{figure}[ht!]
\centering
\includegraphics[width= 5in]{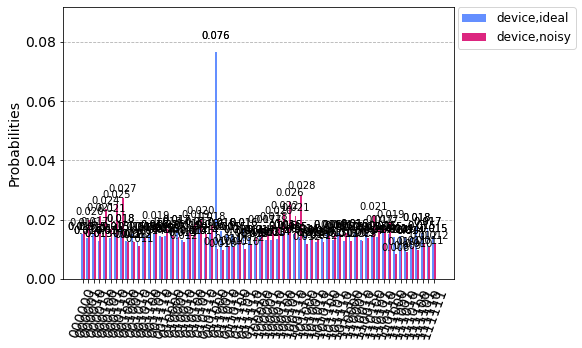}
\caption{Ideal simulation vs Noisy simulation}
\label{compare}
\end{figure}

 As discussed in the background, quantum gates and qubits are noisy and error prone. So, for a realistic result, quantum circuit can be implemented with noise model obtained from real quantum device. Qiskit Aer offers ten standard error models, like Pauli Error, Phase Damping Error, Mixed Unitary Error, Depolarization Error, Reset Error,  Thermal Error where user can customize the error models. In addition, the user can choose whether to apply the error to all qubits or a specific set of qubits. We have executed the quantum circuit of Figure \ref{cliquetriangle} on the remote $ibmq\_qasm\_simulator$, applying a noise model obtained from the available data about $ibmq\_16\_melbourne$ . Executing again the algorithm with the mentioned noise model gives us the histogram in Figure \ref{outnoise}: the peaks are no more clearly distinguishable and the noise caused a loss of information. Figure \ref{compare} shows the comparison between the two histograms, with ideal computation in blue and real (noisy) computation in orange. 

\subsection{Analysis of the Result}
We observe from the implementation results in the real quantum device, that the result has been affected by noise, which is obvious from Figure \ref{outnoise}. Here, we calculate percentage of gate error and decoherence error of the circuit of Figure \ref{cliquetriangle}, while implementing in  $IBMQ\_16\_ Melbourne$. The total number of gates of the circuit in Figure \ref{cliquetriangle} are 230. From the data available about $IBMQ\_16\_ Melbourne$, we can mention the following two parameters:
\begin{itemize}
    \item Mean gate error: $2.14 \times 10^{-3}$
    \item Mean measure error: $2.68 \times 10^{-2}$
\end{itemize}
The probability that at least one gate fails is given by the following
                        \par $P$ (at least one gate fails) \par = $1-P$ (all gate succeed)
                         \par = $1-P$ (1 gate succeed)$^{230}$
                          \par   = $1-(1-2.14 \times$ $10^{-3})^{230}=1-0.61095=0.3890$
                        
This result infers that without accounting for decoherence errors, failure rate of our circuit is 38\%. This is a first important piece of information that gives us an idea of the problem of noise in quantum computation.
Now, talking about the errors due to decoherence, we can try to compute the
time needed by the $ibmq\_16\_melbourne$ to finish the execution of our circuit. The
basis gates set used by the $ibmq\_16\_melbourne$ to realize quantum circuits and
translate more complicated gates into simpler ones is composed by $U1$, $U2$, $U3$
and CNOT.  From public data about the $ibmq\_16\_melbourne$ available at \cite{42} and \cite{43}, we can obtain the required data to calculate the following:
\begin{itemize}
\item Time needed for U1: 0ns;
\item Time needed for U2: 100ns + 20ns = 120ns;
\item Time needed for U3: 100ns + 20ns + 100ns + 20ns = 240ns;
\item Roughly average time needed for CNOT: 100ns+20ns+360ns+20ns+100ns+
20ns + 360ns + 20ns = 1000ns.
\end{itemize}
The real circuit depth is equal to 139. Now, let’s suppose that this circuit depth is composed by 80\% of CNOT gates, 15\% bof $U3$ gates and 5\% of $U2$ gates $i.e.,$ we have 111 CNOT gates, 21 $U3$ gates and 7 $U2$ gates.
\begin{itemize}
    \item  Total time required for CNOT gate is $1000 \times 111=111000 $ns.
    \item  Total time required for $U2$ gate is $7 \times 120= 840$ns.
    \item  Total time required for $U3$ gate is $21 \times 240=5040$ns \\
 \end{itemize}
 So, total time required to execute the circuit is $111000+840+5040 =116880 ns= 116.86$µs.
 Hence, our circuit will take approximately 116.86 µs. As coherence times for Melbourne are $T1$ = 71.5µs and $T2$ = 21.4µs, we can infer that we have a relatively low probability to execute the whole circuit without at least one error due to decoherence.  It’s worth to notice that the circuit of Figure \ref{cliquetriangle} stands for a total number of gates equal to 230 and a circuit depth is 139. For larger circuit total number of gates and circuit depth both will be higher. The Qiskit code for generating the simulation graphs is given in the github. The link is \href{https://github.com/abhaduri77/k-clique-code.git}{https://github.com/abhaduri77/k-clique-code.git}.

\section{Complexity Analysis}
 Proposed combinatorial approach for the synthesis of \textit{k}-clique problem in this paper is based on Grover's algorithm, which gives quadratic speed up than its classical counter part. As the total number of input vertices are $n$ and clique size is \textit{k}, the oracle and the diffusion of Grover’s algorithm needs to be iterated $O(\sqrt{\dfrac{{n \choose k}}{m}})$ to find $m$
 possible cliques with high probability.

\section{Conclusion}

An automated and generalized approach of quantum circuit synthesis for \textit{k}-clique problem using Grover's algorithm with $O(\sqrt{\dfrac{{n \choose k}}{m}})$ iterations for any given undirected and unweighted graph has been proposed in this paper, which gives all cliques of size $k$ as output. We exhibited that the proposed approach of solving $k$-clique problem is cost efficient with respect to qubit and circuit depth, when $k$ is very small with respect to a large graph. Triangle finding problem is demonstrated as an example of $k$-clique problem, when $k=3$ to showcase the effectiveness of our proposed approach of solving $k$-clique problem. The algorithm for \textit{k}-clique problem is further used to solve maximum clique problem in classical-quantum hybrid setting. An automated end-to-end framework for mapping clique problem to any available quantum computer is also designed and proposed. The generated circuits are implemented in NISQ devices and $ibmq\_qasm\_simulator$. From the results, we have observed that the current real quantum computers are affected by some important limitations, like the maximum number of available qubits and different kinds of noise and circuit depth. In future scope of this paper, we shall try to apply some error mitigation scheme that can reduce the error of the output in the real quantum device so that we can have the expected output. We shall also try to reduce the number of gates and ancilla qubit for designing the circuit for clique problem. These will help one to implement clique problem for more larger graphs with large $k$.

\section*{Acknowledgments}
There is no conflict of interest. All the authors contributed equally to this work.

\bibliographystyle{unsrt}  
\bibliography{references}

\begin{thebibliography}{10}

\bibitem{31}
Peter~W. Shor.
\newblock Polynomial-time algorithms for prime factorization and discrete
  logarithms on a quantum computer.
\newblock {\em SIAM Journal on Computing}, 26(5):1484–1509, Oct 1997.

\bibitem{32}
L.~K. Grover.
\newblock A fast quantum mechanical algorithm for database search.
\newblock In {\em Proceedings of the Twenty-eighth Annual ACM Symposium on
  Theory of Computing}, STOC '96, pages 212--219, New York, NY, USA, 1996. ACM.

\bibitem{33}
Frederic {Magniez}, Miklos {Santha}, and Mario {Szegedy}.
\newblock {Quantum Algorithms for the Triangle Problem}.
\newblock {\em arXiv e-prints}, pages quant--ph/0310134, October 2003.

\bibitem{34}
Harry Buhrman and Robert \v{S}palek.
\newblock Quantum verification of matrix products.
\newblock In {\em Proceedings of the Seventeenth Annual ACM-SIAM Symposium on
  Discrete Algorithm}, SODA '06, page 880–889, USA, 2006. Society for
  Industrial and Applied Mathematics.

\bibitem{41}
Richard Karp.
\newblock Reducibility among combinatorial problems.
\newblock volume~40, pages 85--103, 01 1972.

\bibitem{8}
M.~Pavan and M.~Pelillo.
\newblock A new graph-theoretic approach to clustering and segmentation.
\newblock In {\em 2003 IEEE Computer Society Conference on Computer Vision and
  Pattern Recognition, 2003. Proceedings.}, volume~1, pages I--I, 2003.

\bibitem{1}
A.E. Brouwer, J.B. Shearer, N.J.A. Sloane, and W.D. Smith.
\newblock A new table of constant weight codes.
\newblock {\em IEEE Transactions on Information Theory}, 36(6):1334--1380,
  1990.

\bibitem{9}
Vladimir Boginski, Sergiy Butenko, and Panos~M. Pardalos.
\newblock Statistical analysis of financial networks, 2005.

\bibitem{5}
Lizhen Wang, Lihua Zhou, Joan Lu, and Jim Yip.
\newblock An order-clique-based approach for mining maximal co-locations.
\newblock {\em Inf. Sci.}, 179(19):3370–3382, September 2009.

\bibitem{sota}
Sara~Ayman Metwalli, François Le~Gall, and Rodney Van~Meter.
\newblock Finding small and large $k$-clique instances on a quantum computer.
\newblock {\em IEEE Transactions on Quantum Engineering}, 1:1--11, 2020.

\bibitem{2}
Santo Fortunato.
\newblock Community detection in graphs.
\newblock {\em Physics Reports}, 486(3):75--174, 2010.

\bibitem{3}
G.~Palla, I.~Der{\'e}nyi, I.~Farkas, and T.~Vicsek.
\newblock Uncovering the overlapping community structure of complex networks in
  nature and society.
\newblock {\em Nature}, 435:814--818, 2005.

\bibitem{4}
Sercan Sadi, Şule Öğüdücü, and A.~Şima Uyar.
\newblock An efficient community detection method using parallel clique-finding
  ants.
\newblock In {\em IEEE Congress on Evolutionary Computation}, pages 1--7, 2010.

\bibitem{6}
T.~Matsunaga, C.~Yonemori, and E.~Tomita.
\newblock Clique-based data mining for related genes in a biomedical database.
\newblock {\em BMC Bioinformatics}, 10(205), 2009.

\bibitem{7}
R.~E. Bonner.
\newblock On some clustering techniques.
\newblock {\em IBM Journal of Research and Development}, 8(1):22--32, 1964.

\bibitem{27}
IBM.
\newblock Ibmresearch.qiskitsdk0.5.3documentation.
\newblock May 2018.

\bibitem{28}
IBM.
\newblock Quantum computer composer.
\newblock April 2018.

\bibitem{29}
IBM.
\newblock Qiskit - open source quantum information software kit.
\newblock June 2021.

\bibitem{chuang}
M.~A. Nielsen and I.~L. Chuang.
\newblock {\em Quantum Computation and Quantum Information: 10th Anniversary
  Edition}.
\newblock Cambridge University Press, 2010.

\bibitem{PhysRevA.52.3457}
Adriano Barenco, Charles~H. Bennett, Richard Cleve, David~P. DiVincenzo, Norman
  Margolus, Peter Shor, Tycho Sleator, John~A. Smolin, and Harald Weinfurter.
\newblock Elementary gates for quantum computation.
\newblock {\em Phys. Rev. A}, 52:3457--3467, Nov 1995.

\bibitem{Preskill2018quantumcomputingin}
John Preskill.
\newblock Quantum {C}omputing in the {NISQ} era and beyond.
\newblock {\em {Quantum}}, 2:79, August 2018.

\bibitem{45}
Barry~C Sanders.
\newblock {\em How to Build a Quantum Computer}.
\newblock 2399-2891. IOP Publishing, 2017.

\bibitem{43}
IBM~Quantum Experience.
\newblock Ibmq\_16\_melbourne.

\bibitem{17}
M.~Aghaei, Z.~Zukarnain, A.~Mamat, and H.~Zainuddin.
\newblock A hybrid architecture approach for quantum algorithms.
\newblock {\em Journal of Computer Science}, 5:725--731, 2009.

\bibitem{40}
J.~Cirasella.
\newblock Classical and quantum algorithms for finding cycles.
\newblock {\em MSc Thesis}, pages 1--58, 2006.

\bibitem{37}
M.~Szegedy.
\newblock On the quantum query complexity of detecting triangles in graphs.
\newblock {\em arXiv: Quantum Physics}, 2003.

\bibitem{38}
Francois~Le Gall.
\newblock Improved quantum algorithm for triangle finding via combinatorial
  arguments.
\newblock In {\em 2014 IEEE 55th Annual Symposium on Foundations of Computer
  Science}, pages 216--225, 2014.

\bibitem{39}
Fran\c{c}ois Gall and Shogo Nakajima.
\newblock Quantum algorithm for triangle finding in sparse graphs.
\newblock {\em Algorithmica}, 79(3):941–959, November 2017.

\bibitem{15}
Pronaya~Prosun Das and Mozammel H.~A. Khan.
\newblock Solving maximum clique problem using a novel quantum-inspired
  evolutionary algorithm.
\newblock In {\em 2015 International Conference on Electrical Engineering and
  Information Communication Technology (ICEEICT)}, pages 1--6, 2015.

\bibitem{14}
Kuk-Hyun Han and Jong-Hwan Kim.
\newblock Quantum-inspired evolutionary algorithm for a class of combinatorial
  optimization.
\newblock {\em IEEE Transactions on Evolutionary Computation}, 6(6):580--593,
  2002.

\bibitem{16}
Elijah Pelofske, Georg Hahn, and Hristo~N. Djidjev.
\newblock Solving large maximum clique problems on a quantum annealer, 2019.

\bibitem{Acasiete_2020}
F.~Acasiete, F.~P. Agostini, J.~Khatibi Moqadam, and R.~Portugal.
\newblock Implementation of quantum walks on ibm quantum computers.
\newblock {\em Quantum Information Processing}, 19(12), Nov 2020.

\bibitem{zul8342181}
Alwin Zulehner, Alexandru Paler, and Robert Wille.
\newblock Efficient mapping of quantum circuits to the ibm qx architectures.
\newblock In {\em 2018 Design, Automation Test in Europe Conference Exhibition
  (DATE)}, pages 1135--1138, 2018.

\bibitem{li2019tackling}
Gushu Li, Yufei Ding, and Yuan Xie.
\newblock Tackling the qubit mapping problem for nisq-era quantum devices,
  2019.

\bibitem{tan2020optimal}
Bochen Tan and Jason Cong.
\newblock Optimal layout synthesis for quantum computing, 2020.

\bibitem{42}
IBM~Quantum Experience.
\newblock Overview of quantum gates.

\end{thebibliography}

\end{document}